\documentclass[a4paper,11pt]{article}	
\pdfoutput=1 
\usepackage{jheppub} 
\usepackage{tikz}    
\usepackage{amsmath,amssymb,amsthm,amscd,graphicx,amsfonts}
\usepackage{autobreak}
\usepackage{enumerate}
\usepackage{ytableau}
\newcommand{\bitem}{\begin{itemize}}
\newcommand{\eitem}{\end{itemize}}
\newcommand{\be}{\begin{equation}}
\newcommand{\ee}{\end{equation}}
\newcommand{\ba}{\begin{aligned}}
\newcommand{\ea}{\end{aligned}}

\usepackage{enumerate}
\usepackage{ytableau}

\newcommand{\T}{\mathsf{T}}

\usepackage{multirow}

\def\rh{\rho^{\rm F}(\sigma_p)}
\def\SL{SL(2,\IZ)}

\def\chiNSt{\chi_{\rm \widetilde{NS}}} 
\def\chiR{\chi_{\rm R}}

\def\IZ{{\mathbb Z}}

\def\IC{{\mathbb C}}

\newcommand{\tc}{{\tilde c}}

\newcommand{\tilh}{{\tilde h}}
\newcommand{\tm}{{\tilde m}}

\newcommand{\cN}{\mathcal N}
\newcommand{\cF}{\mathcal F}
\newcommand{\cS}{\mathcal S}
\newcommand{\TF}{\mathsf{T}^{\rm F}}

\title{\boldmath Hecke Relations among 2d Fermionic RCFTs}
\author{Kimyeong Lee and Kaiwen Sun}
  
\affiliation{Korea Institute for Advanced Study,
85 Hoegiro, Dongdaemun-gu, Seoul 02455, Korea}

\emailAdd{klee@kias.re.kr}
\emailAdd{ksun@kias.re.kr}

\abstract{Recently, Harvey and Wu proposed a suitable Hecke operator for vector-valued $SL(2,\IZ)$ modular forms to connect the characters of different 2d rational conformal field theories (RCFTs). We generalize such an operator to the 2d fermionic RCFTs and call it \emph{fermionic Hecke operator}. The new Hecke operator naturally maps the Neveu-Schwarz (NS) characters of a fermionic theory to the NS characters of another fermionic theory.  Mathematically, it is the natural Hecke operator on vector-valued $\Gamma_\theta$ modular forms of weight zero. We find it can also be extended to $\rm \widetilde{NS}$ and Ramond (R) sectors by combining the characters of the two sectors together.  
We systematically study the fermionic Hecke relations among 2d fermionic RCFTs with up to five NS characters and find that almost all known supersymmetric RCFTs can be realized as fermionic Hecke images of some simple theories such as supersymmetric minimal models.  We also study the coset relations between fermionic Hecke images with respect to $c=12k$ holomorphic SCFTs.}

\begin{document}

\setcounter{tocdepth}{2}
\maketitle
\flushbottom

\section{Introduction}
In recent years there has been a resurgence of interest in 2d fermionic and supersymmetric conformal field theories. Much progress has been made from various approaches including fermionic modular bootstrap \cite{Bae:2020xzl,Bae:2021mej,Duan:2022kxr}, super modular category (SMC) \cite{Lan:2015sgw,Bruillard:2016yio,Bruillard:2018,Bruillard:2019,Bruillard2019ClassificationOS,Cho:2022kzf}, super vertex operator algebra (SVOA) \cite{duncan05,Creutzig:2017fuk,Johnson-Freyd:2019wgb,Harrison:2020wxl}, topological modular forms \cite{Albert:2022gcs},  fermionization  by non-anomalous $\IZ_2$ symmetry \cite{Tachikawa18,Karch:2019lnn,Runkel:2020zgg,Hsieh:2020uwb,Kulp:2020iet,Bae:2021lvk,Kikuchi:2022jbl}  and so on. In particular, many new fascinating examples of fermionic and supersymmetric CFTs were found which largely broaden our scope beyond the classical constructions in 1980s such as the $\cN=1,2$ supersymmetric minimal models. Not only many WZW models now are shown to have emergent supersymmetry \cite{Bae:2021lvk}, many sporadic groups are also associated to SVOA structures for which a series of renowned examples is the Suzuki chain \cite{Johnson-Freyd:2019wgb}. 

In this work, we introduce a different approach called the \emph{fermionic Hecke operator} which emphasizes more on the fermionic characters and their modularity  of 2d fermionic rational CFTs. The Hecke operator for 2d bosonic RCFTs was introduced by Harvey and Wu \cite{Harvey:2018rdc} and later developed in \cite{Harvey:2019qzs,Wuthesis,Duan:2022ltz}. Such operator exploits the modularity of  vector-valued modular forms and elegantly connects the characters of 2d RCFTs with different central charges. It also reveals new interesting number-theoretic properties of characters as well as nontrivial relations on the modular representations and fusion algebras in the space of  RCFTs. The Hecke operator along with the coset with respect to $c=8k$ holomorphic theories also provides a new paradigm to classify 2d bosonic RCFTs \cite{Duan:2022ltz}. 
It is natural to consider how to generalize such an operator to the fermionic cases and utilize  it to study various aspects of 2d fermionic RCFTs. This question was raised at the end of \cite{Duan:2022ltz}.

As the first step, we find that the most natural setting to define fermionic Hecke operator is in the Neveu-Schwarz (NS) sector of fermionic RCFTs.  The main reason lies in that the NS characters transform to themselves under the $S$ modular action. More precisely, they form a weight-zero vector-valued modular form of a level-two congruence subgroup $\Gamma_\theta$ of $\SL$, which is generated by the $S$ and $T^2$ actions. This setting may be not surprising since the super modular category is also defined by the NS data. We define the fermionic Hecke operator $\TF_p$ on the space of weight-zero $\Gamma_\theta$ vector valued modular forms  such that it naturally maps the NS characters of a fermionic RCFT of central charge $c$ to the NS characters of another fermionic RCFT of central charge $pc$. This resembles the basic property of the bosonic Hecke operator. On the other hand, as we will show later, the $\rm \widetilde{NS}$ and R characters combined together also form a $\Gamma_\theta$ vector-valued modular form, thus the fermionic Hecke operator can be applied on the $(\rm \widetilde{NS},R)$ sector as well.

To make the fermionic Hecke operator $\TF_p$ consistent with the bosonic Hecke operator $\T_p$ defined in \cite{Harvey:2018rdc}, a prerequisite is that fermionization $\cF$ should be commutative with Hecke operation. This means that for a bosonic theory B allowing fermionization, $\TF_p(\cF\textrm{B})$ should be the same with $\cF (\T_p\textrm{B})$. We find that our definition of $\TF_p$ indeed satisfies this nontrivial condition. Note that such commutativity does not diminish the value of fermionic Hecke operator for two reasons. One main reason is that when a (potential) fermionic theory is bootstraped, e.g. from the fermionic modular linear differential equations (MLDEs) for NS characters, or a (potential) object of super modular category is found, it is sometimes not easy to determine precisely what the underlying full bosonic theory is or what the full bosonic characters are. This urges an independent approach to establish maps just by the NS data. Our definition of fermionic Hecke operator successfully realizes this goal, where a key ingredient interestingly is a function very recently proposed in \cite{Cho:2022kzf} in the study of the super modular category. Besides, from the computational aspects, it is also much easier to directly proceed with just some fermionic characters, rather than to first recover the full bosonic characters, perform the bosonic Hecke operator and take fermionization in the end, which sometimes is even impossible.

We summarize some nice examples of fermionic Hecke images we find in the following Table \ref{goodex}, which are also the main results of the current paper. Here we use $SM_{\rm eff}(p',p)$ to denote the effective theory of supersymmetric minimal model $SM(p',p)$. We use $F$ for a single chiral fermion and $\cS$ for supersymmetrization. The $\cS^2$ stands for $\mathcal{N}=2$ supersymmetry. The details will be given later.
\begin{table}[ht]
\def\arraystretch{1.1}
	\centering
	\begin{tabular}{|c|c|c|c|c|c|c|c|c|c|c|c|c|}
		\hline
 Theory	&	 $c$ & $h_{\rm NS}$   & $\TF_p$ & Theory & $\tc$ & $\tilh_{\rm NS}$     \\
		\hline
	\multirow{3}{*}{$F$}  & \multirow{3}{*}{$\frac{1}{2}$} & \multirow{3}{*}{$-$} & $\TF_{31}$  &  $\cF(E_8)_2$ & $\frac{31}{2}$ &  $-$ \\  
  &  &	& $\TF_{35}$  &  $\cF(C_{10})_1$ & $\frac{35}{2}$ &  $-$ \\  
   &  &	& $\TF_{47}$  &  $\mathbb{B}$ & $\frac{47}{2}$ &  $-$ \\  \hline 
$\cF(A_1)_1^2$ & 2 & $-$ & $\TF_7$ & $\cF(E_7)_1^2$ &$14$   &  $-$ \\  \hline 
 $\cF(A_3)_1$ & $3$ &  $-$ & $\TF_{5}$  & $\cF(A_{15})_1$ & $15$ &  $-$   \\  \hline 
\multirow{2}{*}{$SM_{\rm eff}(8,2)$}  & \multirow{2}{*}{$\frac{3}{4}$} & \multirow{2}{*}{$\frac{1}{4} $} & $\TF_3$  & $\cF(A_{1})_6$  & $\frac{9}{4}$ & $\frac{1}{4} $  \\
& & & $\T_{13 }^{\rm F}$ & $\cF(C_{6})_1$  & $\frac{39}{4}$ & $\frac{3}{4} $     \\
		\hline
		$SM(5,3)$ & $\frac{7}{10}$ & $\frac{1}{10}$ & $\TF_{19}$ &  $\cF (E_7)_2$  & $\frac{133}{10}$ &$\frac{1}{10}$ \\  \hline  
$E_8$ inv of $SM_{\rm eff}(60,2)$  & $\frac{7}{5}$ & $\frac15$	 & $\TF_{13}$ & $\cF(D_7)_3$ & $\frac{91}{5}$ & $\frac{11}{10}$ \\  \hline
$\cS A_1$ & $\frac32$ & $\frac14$ & $\TF_{7}$   &  $\cS E_7$ & $\frac{21}{2}$  & $\frac34$ \\  \hline
$\cS^2 A_1$ & $1$ & $(\frac16)_2$ & $\TF_{11}$ &  $\cF(A_{11})_1$ & $11$ & $(\frac{5}{6})_2$  \\ 	\hline
$\cS^2 A_1\otimes F$ & $\frac32$ & $(\frac16)_2$ & $\TF_{13}$ &  $\cF(E_6)_4$ &  $\frac{39}{2}$  & $(\frac{7}{6})_2$ \\		\hline
\multirow{2}{*}{$SM_{\rm eff}(12,2)$} & \multirow{2}{*}{$1$} & \multirow{2}{*}{$\frac16,\frac12$} & $\TF_5$ & $\cF (B_2)_3$ & $5$ &  $\frac13,\frac12 $\\		
 &  &  &  $\TF_7$ & $\cF (C_3)_2$ & $7$ & $\frac12,\frac23$ \\		\hline
{$SM_{\rm eff}(8,2)^2$} & {$\frac32$} & {$(\frac{1}{4})_2,\frac{1}{2}$} & $\TF_5$ &  $\cF (A_3)_4$  & $\frac{15}{2}$ & $ (\frac14)_2,1$ \\
\hline
\multirow{2}{*}{$D_6$ inv of $SM_{\rm eff}(20,2)$} & \multirow{2}{*}{$\frac65$} & \multirow{2}{*}{$\frac{1}{10},(\frac{3}{10})_2$} & $\TF_3$ &  $\cF (A_1)_3^2$ & $\frac{18}{5}$ &  $\frac{3}{10},(\frac25)_2$ \\
& & & $\TF_7$ & $\cF (C_3)_1^2$ & $\frac{42}{5}$ & $\frac{7}{10},(\frac35)_2$ \\\hline
$SM_{\rm eff}(16,2)$ & $\frac{9}{8}$ & $\frac{1}{8},\frac38,\frac{3}{4}$  & $\TF_7$& $\cF (B_3)_3$   & $\frac{63}{8}$   & $\frac14,\frac38,\frac58$  \\\hline
$SM(6,4)$ & $1$ & $\frac{1}{16},\frac16,1$ & $\TF_{11}$ & $\cF(D_6)_2$ & $11$ &  $\frac{11}{16},\frac{5}{6},1 $   \\\hline
inv of $SM(8,6)$ & $\frac54$ &  $(\frac{1}{12})_2,\frac{1}{4},(\frac{5}{6})_2$  & $\TF_{11}$ & $\cF(A_5)_2$ &  $\frac{55}{4}$  &  $ (\frac{7}{12})_2,\frac{3}{4},(\frac{5}{6})_2 $    \\\hline
$D_8$ inv of $SM_{\rm eff}(28,2)$ & $\frac{9}{7}$  & $\frac{1}{14},\frac{3}{14},(\frac37)_2$  & $\TF_5$ & $\cF (D_3)_3$ & $\frac{45}{7}$ & $\frac{5}{14},\frac{9}{14},(\frac47)_2$  \\\hline
$D_{10}$ inv of $SM_{\rm eff}(36,2)$ & $\frac{4}{3}$ & $\!\frac{1}{18},\frac{1}{6},\frac{1}{3},(\frac{5}{9})_2\!$ & $\TF_7$ &  $\cF (D_4)_3$ &  $\frac{28}{3}$ & $\frac{7}{18},(\frac89)_2,\frac56,\frac23$ \\\hline
		\end{tabular}
			\caption{Examples of fermionic Hecke relations. The inv is short for non-diagonal $\Gamma_\theta$ modular invariant. The $h_{\rm NS}$ are the weights of non-vacuum NS primaries with degeneracy marked. The $\mathbb{B}$ stands for Baby Monster SVOA \cite{hoehn}.}
			\label{goodex}
		\end{table}

The bosonic Hecke operator is closely related to the cosets with respect to holomorphic CFTs \cite{Harvey:2018rdc,Harvey:2019qzs,Wuthesis}. These special CFTs are well-known to carry central charge $8k,k\in \IZ$ and a single bosonic character related to the Klein $J$ function. The coset relations with respect to holomorphic CFTs along with the implication in MLDEs were nicely discussed in \cite{Gaberdiel:2016zke}. In \cite{Harvey:2018rdc}, it was realized that many Hecke images can pair together to form a (potential) $c=8k$ holomorphic theory, which suggests that the characters of a pair of Hecke images satisfy a simple bilinear relation. For example, the bilinear combination of the characters of a $c=24$ pair of Hecke images usually gives $J(\tau)+N$, where $N$ is the sum of the number of spin-1 currents for the two Hecke images. In the cases where both Hecke images are 
physical, this gives the character $J(\tau)+N$ of $c=24$ holomorphic CFTs classified by Schellekens \cite{Schellekens:1992db}. We find these nice properties can be generalized to the fermionic cases, where we are interested in the pairs of fermionic Hecke images forming $c=12k$ holomorphic SCFTs which have one single NS character. The bilinear combination of such a pair of fermionic Hecke images produces a degree $k$ polynomial of a function $K(\tau)=(\theta_3/\eta)^{12}-24$. 
The simplest case, i.e. the $c=12$ holomorphic SCFTs have been classified by Creutzig, Duncan and Riedler in 2017 \cite{Creutzig:2017fuk} which contain only three cases: supersymmetric $E_{8,1}$ theory ${\cS E_8}$, Conway SCFT \cite{duncan05} and the theory of 24 free chiral fermions $F_{24}$. We will show that indeed the characters of a $c=12$ pair of physical fermionic Hecke images always have bilinear combination equal to $K(\tau),K(\tau)+8$ or $K(\tau)+24$, which are the single NS character of Conway SCFT, ${\cS E_8}$ or $F_{24}$ respectively. 
 
Hecke operator is also known to have deep connection with the modular tensor category (MTC) \cite{Harvey:2019qzs}. The data of MTC is defined by central charge modulo 8 and conformal weights modulo 1. It was found in \cite{Harvey:2019qzs} that Hecke operator can induce Galois conjugate in MTC. The classification results with a small number of characters in \cite{Duan:2022ltz} also suggest that Hecke operator along with $c=8k$ cosets should be able to produce all objects in MTC. We notice that this paradigm can be generalized to fermionic Hecke operator and super modular category in which the SMC data is defined by central charge and NS weights both modulo $1/2$. In fact, we conjecture that the fermionic Hecke operator and some proper cosets can uniformly generate all objects in SMC. We leave the discussion on relation between fermionic Hecke operator and SMC in future work and focus on the aspect of characters in the current one.

This paper is organized as follows. In Section \ref{sec:review}, we give an overview on the 2d fermionic and supersymmetric RCFTs. We will review many known examples and the concept of cosets with respect to $c=12$ holomorphic SCFTs. In Section \ref{sec:hecke}, we introduce our new approach -- fermionic Hecke operator and discuss its main properties. In particular, we will discuss the simplest yet still nontrivial example of free chiral fermions.  In Sections \ref{sec:2chi}, \ref{sec:3chi}, \ref{sec:4chi}, \ref{sec:5chi}, we will discuss the fermionic Hecke operation on some small fermionic theories with $2,3,4,5$ NS characters (modulo degeneracy) respectively. We will show that almost all known supersymmetric RCFTs can be realized as some fermionic Hecke images. In Section \ref{sec:R}, we briefly discuss the fermionic Hecke operation for $\rm \widetilde{NS}$ and R sectors. In Section \ref{sec:outlook}, we conclude and raise some questions for future work. Following our previous paper \cite{Duan:2022ltz}, we use $\TF_p,\star$ to represent fermionic Hecke images with  NS quasi-characters which means there are negative Fourier coefficients. These images usually have coefficient $-1$ for the NS vacuum state, thus are unphysical theories. Besides, we adopt the convention of affine Lie algebras as in SageMath \cite{sage}.

\section{Basics on 2d fermionic RCFTs}\label{sec:review}
\subsection{Basics of fermionic RCFTs}\label{sec:basic}
2d RCFTs have rational central charge, rational conformal weights and a finite number of conformal primaries. For bosonic RCFTs, the torus partition function can be written as a sesquilinear combination of characters
\be
Z(\tau,\bar{\tau})=\sum_{i,j=0}^{d-1} M_{ij} \chi_i(\tau)\overline{\chi}_j(\bar{\tau}),\qquad  M_{ij} \in\mathbb{N}_{\ge0}.
\ee
The characters $\chi_i(\tau)$ together form a $d$-dimensional vector valued modular form of $\SL$ such that the torus partition function is modular invariant. More precisely, there exists a $d$-dimensional representation $\rho:\SL\to GL(d,\IC)$ such that for any $\gamma\in\SL$, 
\be
\chi_i(\gamma \tau)=\sum_{j}\rho(\gamma)_{ij}\chi_j(\tau).
\ee
For the two generators of $\SL$,
\be
T=\left(
\begin{array}{cc}
 1 & 1 \\
 0 & 1 \\
\end{array}
\right) ,\quad  S=\left(
\begin{array}{cc}
 0 & -1 \\
 1 & 0 \\
\end{array}
\right),
\ee
the $\rho(T)$ and $\rho(S)$ are usually called the $T$ and $S$ matrices of the bosonic RCFT.  The $T$ and $S$ matrices  satisfy some constraints. For example, $C=\rho(S)^2=(\rho(T)\rho(S))^3$ gives the charge conjugation matrix of the theory.

In the presence of fermions, it is necessary to specify the boundary conditions along the two cycles of the torus. There are either periodic, i.e., Ramond (R) condition or anti-periodic, i.e., Neveu-Schwarz (NS) condition for each cycle. Apparently, there exist four possible combinations of boundary conditions along the two cycles: $(\rm NS,NS)$, $(\rm R,NS)$, $(\rm NS,R)$, $(\rm R,R)$. It is traditional to call them the NS, $\rm \widetilde{NS}$, R, $\rm \widetilde{R}$ sectors respectively. These are also called the spin structures. In each sector, there is torus partition function defined by the trace over Hamiltonians as
\be
\ba
Z_{\mathrm{NS}}(\tau,\bar{\tau} ) &= \mathrm{Tr}_{\mathrm{NS}}(q^{L_0-\frac{c}{24}}\bar{q}^{\bar{L}_0-\frac{c}{24}}),\qquad  Z_{\widetilde{\mathrm{NS}}}(\tau,\bar{\tau} ) = \mathrm{Tr}_{\mathrm{NS}}(q^{L_0-\frac{1}{2}}\bar{q}^{\bar{L}_0-\frac{c}{24}}(-1)^F), \\
   Z_{\mathrm{R}}(\tau,\bar{\tau} ) &= \mathrm{Tr}_{\mathrm{R}}(q^{L_0-\frac{1}{2}}\bar{q}^{\bar{L}_0-\frac{c}{24}}),\qquad\ \ \ \ Z_{\widetilde{\mathrm{R}}}(\tau,\bar{\tau} ) = \mathrm{Tr}_{\mathrm{R}}(q^{L_0-\frac{1}{2}}\bar{q}^{\bar{L}_0-\frac{c}{24}}(-1)^F).
\ea
\ee
Each of these partition functions can be written as a sesquilinear combination of the NS, $\rm \widetilde{NS}$, R, $\rm \widetilde{R}$ characters respectively. For example,
\be
Z_{\mathrm{NS}}(\tau,\bar{\tau} ) =\sum_{i,j=0}^{n-1} M_{ij}^{\rm NS} \chi_i^{\rm NS}(\tau)\overline{\chi}_j^{\rm NS}(\bar{\tau}),\qquad  M_{ij}^{\rm NS} \in\mathbb{N}_{\ge0}  .
\ee
Unlike the bosonic case, the four sectors of fermionic theory do not necessarily transform to the same sector under $\SL$. By careful analysis on the boundary conditions, it can be found that the NS characters transform to themselves under $S$ generator, and to $\rm \widetilde{NS}$ characters under $T$ generator. 
We summarize the transformation among different sectors by the following formula:
\be
T\textrm{   :  }\ 
\left(
\begin{array}{c}
\chi_{\rm NS} \\
\chi_{\rm \widetilde{NS}} \\
\chi_{\rm R}  \\
\chi_{\rm \widetilde{R}} \\
\end{array}
\right)(\tau+1)=\left(
\begin{array}{cccc}
 0 & 1 & 0 & 0 \\
 1 & 0 & 0 & 0 \\
 0 & 0 & 1 & 0 \\
 0 & 0 & 0 & 1 \\
\end{array}
\right)\cdot\left(
\begin{array}{c}
\chi_{\rm NS} \\
\chi_{\rm \widetilde{NS}} \\
\chi_{\rm R}  \\
\chi_{\rm \widetilde{R}} \\
\end{array}
\right)(\tau).
\ee
\be
S\textrm{    :  }\ 
\left(
\begin{array}{c}
\chi_{\rm NS} \\
\chi_{\rm \widetilde{NS}} \\
\chi_{\rm R}  \\
\chi_{\rm \widetilde{R}} \\
\end{array}
\right)\left(-\frac{1}{\tau}\right)=\left(
\begin{array}{cccc}
 1 & 0 & 0 & 0 \\
 0 & 0 & 1 & 0 \\
 0 & 1 & 0 & 0 \\
 0 & 0 & 0 & 1 \\
\end{array}
\right)\cdot\left(
\begin{array}{c}
\chi_{\rm NS} \\
\chi_{\rm \widetilde{NS}} \\
\chi_{\rm R}  \\
\chi_{\rm \widetilde{R}} \\
\end{array}
\right)(\tau).
\ee
With the above $T$ and $S$ transformations, the full fermionic partition function 
\be
Z_{\rm F}=\frac{1}{2}\left(Z_{\rm NS}+Z_{\rm \widetilde{NS}}+Z_{\rm R}+Z_{\rm \widetilde{R}}\right)
\ee
is still modular invariant under $SL(2,\IZ)$. Besides, the above transformations imply that the number of NS, $\rm \widetilde{NS}$ and R characters are also the same, while the number of $\rm \widetilde{R}$ characters is independent.

Although NS, $\rm \widetilde{NS}$ and R characters are not modular under $\SL$, 
the analysis in \cite{Bae:2020xzl} showed that 
they are  vector valued modular forms of the following level-two congruence subgroups  respectively
\be
\ba 
\Gamma_\theta&=  \Big\{ \gamma\in \SL \big| \gamma\equiv \left({_1 \ _0 \atop ^0 \ ^1}\right) \ {\rm or}\   \left({_0 \ _1  \atop ^1 \ ^0} \right) \, {\rm mod} \ 2 \Big\}, \\
    \Gamma^0(2)&=  \Big\{ \gamma\in \SL \big| \gamma\equiv \left({_* \ _0 \atop ^* \ ^ * }\right)  \   {\rm mod} \ 2\Big\},\\
    \Gamma_0(2)&=   \Big\{ \gamma\in \SL \big| \gamma\equiv \left({_* \ _* \atop ^0 \ ^*}\right)   \   {\rm mod}\ 2 \Big\}.
\ea
\ee
For example, suppose there are $n$ number of NS characters, then there exists a $n$-dimensional representation $\rho^{\rm F}:\Gamma_\theta\to GL(n,\IC)$ such that for any $\gamma\in\Gamma_\theta$, 
\be
\chi_i^{\rm NS}(\gamma \tau)=\sum_{j}\rho^{\rm F}(\gamma)_{ij}\chi_j^{\rm NS}(\gamma \tau).
\ee
It is easy to see that the $\Gamma_\theta$ congruence subgroup is generated by $S$ and $T^2$ of $\SL$. Therefore, it is convenient to call $\rho^{\rm F}(S)$ and $\rho^{\rm F}(T^2)$ as the $S$ and $T^2$ matrices of (the NS characters of) a fermionic RCFT. We sometimes omit the $\rho^{\rm F}$ when there is no room for confusion. From now on, we focus on the $S$ and $T^2$ matrices of the NS sector. Such matrices are known to satisfy the following constraints
\be
\rho^{\rm F}(S)^4={\rm Id},\qquad \rho^{\rm F}(S)^2\rho^{\rm F}(T^2)=\rho^{\rm F}(T^2)\rho^{\rm F}(S)^2.
\ee
Using $\rho^{\rm F}(S)$ one can define the fusion rules of NS primaries inside the NS sector by the usual Verlinde formula \cite{Verlinde:1988sn}. 

The NS, $\rm \widetilde{NS}$ and R characters in general have the following kinds of Fourier expansion
\be
\ba
\chi_{\rm NS}&=q^{-\frac{c}{24}+h_{\rm NS}} (m_0+m_{1/2}q^{1/2}+m_1q+m_{3/2}q^{3/2}+\dots),\\
\chi_{\rm \widetilde{NS}}&=q^{-\frac{c}{24}+h_{\rm \widetilde{NS}}} (m_0-m_{1/2}q^{1/2}+m_1q-m_{3/2}q^{3/2}+\dots),\\
\chi_{\rm R}&=q^{-\frac{c}{24}+h_{\rm R}} (s_0+s_1q+s_2q^2+\dots).
\ea
\ee
All coefficients $m_i$ and $s_i$ should be non-negative integers and $m_0$ for the NS vacuum should be 1. The $m_{1/2}$ measures the number of free fermions. The $k$ number of NS characters $\chi_i^{\rm NS}$ in general satisfy an order $k$ fermionic MLDE of $\Gamma_\theta$, while the $\chi_i^{\rm \widetilde{NS}}$ and $\chi_i^{\rm R}$ satisfy fermionic MLDEs of $ \Gamma^0(2)$ and $ \Gamma_0(2)$ respectively \cite{Bae:2020xzl}. The three different fermionic MLDEs are equivalent by modular transformations. In the current paper, we will be mainly interested in the NS characters. As a final remark, the $\rm \widetilde{R}$ characters are isolated, and do not have modular connection with NS, $\rm \widetilde{NS}$ and R characters. It is known for supersymmetric RCFTs, there is just one single R character and it is a constant.

\subsection{Examples of fermionic RCFTs}\label{sec:ex}
A large class of $\mathcal{N}=1$ supersymmetric RCFTs are the \emph{supersymmetric minimal models} $SM(p',p)$ for $2\le p\le p'-2$ , $p'-p\in 2\IZ$ and $\gcd(\frac{p'-p}{2},p)=1$ \cite{Friedan:1983xq,Friedan:1984rv,Bershadsky:1985dq}. They have only super Virasoro symmetry and are also called the minimal SCFTs. The central of  $SM(p',p)$ is given by
\be
c=\frac{3}{2}\Big(1-\frac{2(p'-p)^2}{pp'}\Big)
\ee
and the fermionic conformal weights are 
\be
h_{r,s}=\frac{(sp-p'r)^2-(p'-p)^2}{8pp'}+\frac{2\epsilon_{r-s}-1}{16},\qquad \epsilon_a =
\begin{cases}
  \frac12\quad  a\in 2\IZ ,\\
  1 \quad a\in 2\IZ+1.
  \end{cases}
\ee
Here $r=1,2,\ldots,p-1$ and $s=1,2,\ldots,p'-1$. The $h_{r,s}$ with $r-s\in 2\IZ$ belong to the NS sector, while those with $r-s\in 2\IZ+1$ belong to the R sector. Owing to the symmetry $(r,s)\leftrightarrow(p-r,p'-s)$, we only need to consider the $sp\le p'r$ part. Denote $(q)_\infty=\prod_{n=1}^\infty(1-q^n)$.
The fermionic characters of $SM(p',p)$ are given by \cite{Goddard:1986ee}
\be\label{chisum}
\ba
\chi_{r,s}(q) = \chi_{p-r,p'-s}(q) =
 \frac{(-q^{\epsilon_{r-s}})_\infty}{ (q)_\infty}\sum_{\ell\in \IZ} \left( q^{\ell(\ell pp'+rp'-sp)/2} -q^{(\ell p+r)(\ell p'+s)/2}  \right).
\ea
\ee

It is well-known that only for $p'=p+2$, the $SM(p',p)$ is unitary \cite{Friedan:1983xq,Friedan:1984rv,Bershadsky:1985dq}. The non-unitary supersymmetric minimal models and their non-diagonal modular invariants to our knowledge are less studied. Most of them have negative central charges and negative fermionic weights. Mimic the bosonic  minimal models and their effective theories\footnote{For the picture of effective theory of bosonic non-unitary RCFTs, we refer to a good discussion in \cite[Section 3]{Harvey:2019qzs}. For example, the non-unitary Lee-Yang 
minimal model $M(5,2)$ has central charge $c=-\frac{22}{5}$ and non-vacuum conformal weight $h=-\frac15$. One can exchange the notion of vacuum and non-vacuum primaries by shifting $c$ and $h$ simultaneously while keeping $-\frac{c}{24}+h$ invariant. It is easy to see after such shifting, the central charge becomes $c_{\rm eff}=\frac25$, while non-vacuum weight becomes $h_{\rm eff}=\frac15$. This is called the effective Lee-Yang theory. In general, this phenomenon is related to the Galois shuffle  property introduced in \cite{Gannon:2003de}.}, we define the \emph{effective supersymmetric minimal models} $SM_{\rm eff}(p',p)$ for the non-unitary cases such that they have the positive effective central charge
\be
c_{\rm eff}= \frac{3}{2}\Big(1-\frac{8}{pp'}\Big).
\ee
and non-negative effective weights
\be
h_{r,s}^{\rm eff}=\frac{(sp-p'r)^2-4}{8pp'}+\frac{2\epsilon_{r-s}-1}{16} .
\ee

Mimic the level $k$ Lee-Yang model defined in our previous paper \cite{Duan:2022ltz},
we introduce the notion of \emph{level $k$ supersymmetric Lee-Yang models} defined by
\be
(SLY)_k:=SM_{\rm eff}(2,4k+4).
\ee
The series of non-unitary $SM(2,4k+4)$  models was extensively studied in the 1990s, see for example in \cite{Schoutens:1990vb,Melzer:1994qp,Berkovich:1995nx,Berkovich:1995yg}.
They have effective central charge
\be
c=\frac{3}{2}\Big(1-\frac{1}{1+k}\Big)=\frac{3k}{2(1+k)}
\ee
and $k+1$ NS characters with effective weights
\be
h_i^{\rm NS}=\frac{i(i+1)}{4(k+1)},\qquad i=0,1,\dots,k.
\ee
The NS characters have the following infinite product expression \cite{Melzer:1994qp} up to $q^{-c/24+h_i}$:
\be\label{chiprod}
\ba
\chi_{i}(q)= 
  \prod_{n=1}^\infty \big(1-q^{n/2}\big)^{-1},\quad {n\not\equiv 2\,{\rm mod }\,4 ,\textrm{ and } 
   n\not\equiv 0,\pm (2k+1-2i)\,{\rm mod }\,(4k+4)}.
\ea
\ee
The $S$-matrix for the NS characters $\chi_{i}$ can be determined as
\be
S_{ij}=\sqrt{\frac{2}{k+1}}\cos\left(\frac{\pi (2i+1)(2j+1)}{4k+4}\right),\qquad i,j=0,1,\dots,k.
\ee

Similar to the bosonic minimal models, the $\mathcal{N}=1$ minimal models also have interesting and rich modular invariants. The modular invariants of unitary $SM(p+2,p)$ models have been classified by \cite{Kastor:1986ig,Matsuo:1986vc,Cappelli:1986ed}. The modular invariants of non-unitary $\mathcal{N}=1$ minimal models to our knowledge are less studied. In this paper, we find several non-diagonal $\Gamma_\theta$ modular invariants of non-unitary $\mathcal{N}=1$ minimal models which serve as good input to study fermionic Hecke operations. Besides, some special $\mathcal{N}=1$ minimal models can be realized as fermionization of bosonic minimal models including the famous examples $SM(5,3)\equiv \cF M(5,4)$ and $SM(8,2)\equiv \cF M(8,3)$. A less known example found in \cite{Melzer:1994qp} is that $SM(7,3)$ can be realized as the fermionization of the $E_6$ modular invariant of $M(12,7)$. We will encounter all these examples in the main context.

There also exist a series of unitary $\mathcal{N}=2$ minimal models for $k=1,2,3,\dots$, which comes from the coset construction of $\mathcal{N}=2$ superconformal algebras by $SU(2)$ affine Lie algebras \cite{DiVecchia:1986fwg,Kazama:1988qp}. We denote these theories as $\cS^2A_k$. They have central charge $c=3k/(k+2)$ and NS weights $h_{ab}^{\rm NS}=(ab-1/4)/(k+2)$ with $a,b\in\IZ+1/2$ and $0<a,b,(a+b)<k+1$. In the current paper, we will encounter the $A_1$ case with $c=1$, which in fact can be realized as a non-diagonal modular invariant of $SM(6,4)$ theory. For the NS character formulas of $\cS^2A_k$, we refer to e.g. \cite[Equation (5.41)]{Bae:2020xzl}.

Another large class of $\mathcal{N}=1$ RCFTs are the supersymmetric ADE WZW models of level 1. For $G=$ ADE, we can couple the bosonic WZW model $(G)_1$ with $\textrm{rank}(G)$ number of free chiral fermions to produce obviously supersymmetric theories. We denote these models from the supersymmetrization of   WZW $(G)_1$ as $\cS G$ theories. Apparently the central charge of $\cS G$ is just $3/2$ times the central charge of WZW $(G)_1$, and the NS weights are the same with the bosonic conformal weights of WZW $(G)_1$. The $S$ matrix of $\cS G$ is also identical to the one of WZW $(G)_1$. In the analysis of second order $\Gamma_\theta$ MLDEs \cite{Bae:2020xzl}, the $\cS G$ theories for $G=A_1,A_2,D_4,E_6,E_7$ already appeared as solutions. We notice that the non-vacuum NS weights of these theories
$h_{\rm NS}=\frac14,\frac13,\frac12,\frac23,\frac34$ are just the entries not equal to 0 or 1 in the Farey sequence of order 4. This resembles the observation in \cite{Harvey:2018rdc} that the non-vacuum conformal weights of the solutions of bosonic second order MLDEs \cite{Mathur:1988na} are the entries not equal to 0 or 1 in the Farey sequence of order 5.

In the above, we have been reviewing the examples of supersymmetric RCFTs. There are of course many fermionic RCFTs that are not supersymmetric. For example, a series of renowned fermionic RCFTs comes from the fermionization of WZW $(A_1)_{4k+2}$ models, see e.g. \cite{Bruillard:2016yio}. The $\cF (A_1)_{4k+2}$ theory has $k+1$ number of NS, $\rm \widetilde{NS}$ and R characters for each sector and $k$ number of $\rm \widetilde{R}$ character. It is known only for $\cF (A_1)_{6}$, the fermionic theory becomes supersymmetric. The fermionic characters and the affine $(A_1)_{6}$ characters have the following well-known relations:
\be\label{SU26allchi}
\ba
\chi^{\rm NS}_0&=\chi^{(A_1)_6}_0+\chi^{(A_1)_6}_{3/2},\qquad \chi^{\rm NS}_{1/4}=\chi^{(A_1)_6}_{1/4}+\chi^{(A_1)_6}_{3/4},\\
\chi^{\rm \widetilde{NS}}_0&=\chi^{(A_1)_6}_0-\chi^{(A_1)_6}_{3/2},\qquad \chi^{\rm \widetilde{NS}}_{1/4}=\chi^{(A_1)_6}_{1/4}-\chi^{(A_1)_6}_{3/4},\\
\chi^{\rm R}_{3/32}&=\chi^{(A_1)_6}_{3/32}+\chi^{(A_1)_6}_{35/32},\qquad \chi^{\rm R}_{15/32}=\sqrt{2}\chi^{(A_1)_6}_{15/32},\\
\chi^{\rm \widetilde{R}}_{3/32}&=\chi^{(A_1)_6}_{3/32}-\chi^{(A_1)_6}_{35/32}=2,
\ea
\ee
where the last constant equality is the result of Macdonald identity. 
There exist many more supersymmetric RCFTs that come from the fermionization of WZW models. See good summaries in \cite{Johnson-Freyd:2019wgb}, \cite[Table 1]{Bae:2021lvk} and \cite{Bae:2020xzl,Bae:2021mej}.  In a modern viewpoint, these come from the orbifold of a non-anomalous $\IZ_2$ symmetry and the generalized Jordan-Wigner transformation \cite{Karch:2019lnn,Hsieh:2020uwb}. We will show later that almost all known WZW models with emergent supersymmetry can be realized as fermionic Hecke images.

\subsection{Cosets with respect to $c=12k$ holomorphic SCFTs}\label{sec:coset}
Holomorphic 2d SCFTs or the so-called self-dual SVOAs with central charge $c=12$ are the natural supersymmetric analogies of holomorphic CFTs with $c=24$. Similar to the renowned Schellekens' list \cite{Schellekens:1992db} of the 71 holomorphic CFTs with $c=24$, there is also a complete classification of $c=12$ holomorphic SCFTs. It has been proved in \cite{Creutzig:2017fuk} that there are only three cases of such holomorphic SCFTs: supersymmetric WZW $E_{8,1}$ theory ${\cS E_8}$, Conway SCFT \cite{duncan05} and the theory of 24 chiral fermions $F_{24}$. Furthermore, the $F_{24}$ allows eight possible affine Lie algebra structures also called $\cN=1$ structures \cite{Harrison:2020wxl}:
\be\nonumber
(A_1)_{2}^8,\ (A_2)_{3}^3,\ (A_4)_{5},\ (A_3)_{4}(A_1)_{2}^3,\ (B_2)_{3}(G_2)_{4},\ (B_2)_{3}(A_2)_{3}(A_1)_{2}^2,\ (B_3)_{5}(A_1)_{2},\ (C_3)_4(A_1)_2.
\ee
Interestingly, these eight affine Lie algebra structures also appeared earlier in \cite{DW} in the study of Borcherds products and theta blocks. 
It is worthwhile to point out that all these $c=12$ holomorphic SCFTs have further hyperbolic structures, i.e., Borcherds-Kac-Moody superalgebras. For example, the ${\cS E_8}$ has hyperbolization called the fake Monster superalgebra, describing the physical states of 10-dimensinal superstring moving on torus \cite{Scheithauer:1999rp}.  
The BKM superalgebra with Conway symmetry was constructed in \cite{Harrison:2018joy}.
The BKM superalgebras associated to $F_{24}$ were constructed in \cite{Harrison:2020wxl}. The eight affine Lie algebras in $F_{24}$ and the 69 ones in the  Schellekens' list \cite{Schellekens:1992db} have an uniform description as the hyperbolization of affine Lie algebras in \cite{SWW}.

As we mentioned earlier, a holomorphic SCFT with $c=12$ has one single NS character $K(\tau)+n $ where $K(\tau)$ is defined by
\be\label{Ktau}
K(\tau)=(\theta_3/\eta)^{12}-24=q^{-1/2}+276 q^{1/2}+2048 q+11202 q^{3/2}+49152 q^2+\dots.
\ee
The $n=0$ case corresponds to Conway SCFT, while $n=8$ and $n=24$ correspond to ${\cS E_8}$ and $F_{24}$. More generally, a holomorphic SCFT with $c=12k$ has one single NS character that can be written as a degree $k$ polynomial $P_k(K(\tau))$. The $\rm \widetilde{NS}$ and R characters can easily be obtained from the modular transformation of $K(\tau)$, while the $\rm \widetilde{R}$ character is also some constant.

We are interested in the cosets with respect to $c=12k$ holomorphic SCFTs. They are completely parallel to the bosonic generalized cosets discussed in \cite{Gaberdiel:2016zke}, thus we only state the results. Consider a formal coset $\mathcal{C} = \mathcal{G}/\mathcal{H}$ where $\mathcal{G}$ is a $c=12k$ holomorphic SCFT, and $\mathcal{H}$ is a fermionic sub-theory. Then $\mathcal{C}$ can be defined by the chiral algebra generators which have trivial OPE with all those in $\mathcal{H}$ and it is automatically a fermionic theory. The central charge obviously satisfies $c_{\mathcal{C}} = c_\mathcal{G} - c_\mathcal{H}$. More importantly, the weights of non-vacuum NS primaries should satisfy
\be
h^\mathcal{H}_i + h^\mathcal{C}_i = n_i, \quad \text{with}\ 2n_i \in \mathbb{N} .
\ee
The main feature concerning us is the bilinear relation of the NS characters of a $c=12$ pair of two fermionic theories. Given the classification in \cite{Creutzig:2017fuk}, any such pair of physical theories should have
\be\label{Kn}
\chi^{(c)}_{\rm NS}\cdot \chi^{(12-c)}_{\rm NS}=K(\tau)+n,\qquad n=0,8,24.
\ee
This equation is actually a very strong constraint. We will show later that some $c=12$ pairs of fermionic Hecke images produce bilinear relation $ K(\tau)+n$ with $n$ different from $0,8,24$. In those cases, the $c=12$ pair does not correspond to a consistent holomorphic SCFT. On the other hand, we also find many good pairs of fermionic Hecke images indeed producing the admissible $K(\tau)+n$. For example, $\cS A_1$ and its $\TF_7$ image $\cS E_7$ and pair together to produce ${\cS E_8}$. More generally, one can consider holomorphic SCFTs with central charge $c=12k$. The bilinear relation of a $c=12k$ pair of fermionic theories can be written as
\be
\chi^{(c)}_{\rm NS}\cdot \chi^{(12k-c)}_{\rm NS}=P_k(K(\tau)).
\ee
For $k>1$, to our knowledge there is no classification of $c=12k$ holomorphic SCFTs yet. An interesting constraint was recently proposed in \cite{Albert:2022gcs} from the viewpoint of topological modular forms. We expect pairs of fermionic Hecke images can produce many potential holomorphic SCFTs.

\section{Hecke relations}\label{sec:hecke}
\subsection{Bosonic Hecke relations}
To introduce fermionic Hecke operator, we first briefly review the definition of bosonic Hecke operator given by Harvey and Wu \cite{Harvey:2018rdc}. Consider an arbitrary bosonic RCFT with central charge $c$, conformal weights $h_i$ and characters $\chi_i$. Denote the least common denominator of $h_i-c/24$ as $N$, which is called the \emph{conductor} of the theory. Apparently, an equivalent definition is the smallest number $N$ such that $\rho(T)^N=\rm Id$. It was proved by Bantay \cite{Bantay:2001ni} that  each character $\chi_i$ itself is invariant under $\tau\to \gamma\tau$ for any $\gamma\in\Gamma(N)$  defined as
\be
\Gamma(N) =  \left \{ \left({_a \ _b \atop ^c \ ^d}\right) \in SL(2,\IZ) \big| \, a \equiv d \equiv 1\! \pmod{N}, ~b \equiv c \equiv 0\! \pmod{N} \right \} \, .
\ee
Clearly, the congruence subgroup $\Gamma(N)$ is the kernel of the canonical mod $N$ map $\mu_N: \SL\to SL(2,\mathbb{Z}_N)$. 

For a $\SL$ modular form $f(\tau)$ of weight $0$, the Hecke operator $\T_p$ for prime number $p$ is defined by
\be
(\mathsf{T}_p f)(\tau) :=p^{-1}\sum_{a,d>0,ad=p} \sum_{b\,(\textrm{mod }d)}f\Big(\frac{a\tau+b}{d}\Big).
\ee
For modular forms of $\Gamma(N)$, one has to take into consideration the nature of vector-valued modular forms as well. The proper generalization of the Hecke operator to such circumstances was found by Harvey and Wu. 
Denote $\bar{p}$ as the multiplicative inverse of $p$ modulo $N$ and $\sigma_p$ as $\mu_N^{-1}\mathrm{diag}(\bar{p},p)$. 
Then Hecke operator $\mathsf{T}_p$ acts on $f_i(\tau)$ as
\be\label{Tpbose}
(\mathsf{T}_p f)_i(\tau) : = \sum_j \rho_{ij}(\sigma_p) f_j(p \tau) +\sum_{b=0}^{p-1} f_i\Big(\frac{\tau+bN}{p}\Big) .
\ee
This definition does not preserve $\rho(S)$ and $\rho(T)$, thus is not technically an automorphism. However, this is also an advantage as it releases more possibilities which are encoded in the  matrix $\rho(\sigma_p)$.  We will call each $\rho(\sigma_p)$ as a \emph{Hecke class}. All Hecke classes form a finite abelian group related to the quadratic residue modulo $N$. In practice, the transfer matrix $\rho(\sigma_p)$ can be computed by the $S$ and $T$ matrices as
\be
\rho(\sigma_p)=\rho(T^{\bar p} S^{-1}T^{p}S T^{\bar p} S) .
\ee
The Hecke operator  can also equivalently be defined by the map of Fourier coefficients. Suppose 
$f_i(\tau)=\sum_n a_i(n)\,q^\frac{n}{N}$ and $(\T_p f_i)(\tau)=\sum_{n} a^{(p)}_i(n)\,q^\frac{n}{N}$. Then the Hecke operator \eqref{Tpbose} gives the map
\begin{align}\label{Fouriermap}
\begin{split}
a^{(p)}_{i}(n) =
			\left\{
			\begin{array}{ll}
			p\, a_{i}(pn), &\ p \nmid n , \\
			p\, a_{i}(pn)+\sum_{j} \rho_{ij}(\sigma_p) a_j
			\big( \frac{n}{p}\big), &\ p \ | \ n.
			\end{array}
			\right.
		\end{split}
\end{align}
The $p$ divisibility of this kind of series $a^{(p)}_{i}$ is called the \emph{mod $p$ property}. The Hecke operator \eqref{Tpbose} can be further generalized to non-prime $p$ with $\text{gcd}(p,N)=1$ by
\begin{align}
		\begin{split}
\left\{
			\begin{array}{ll}
			\T_{rs}=\T_r\circ \T_s,&\ \text{gcd}(r,s)=1,  \\
			\T_{p^{n+1}}= \T_{p}\circ\T_{p^{n}}-p\sigma_p\circ\T_{p^{n-1}},&\ p\text{ prime.}
			\end{array}
			\right.
		\end{split}
	\end{align}
The $S$ and $T$ matrices of the Hecke image $\T_p$ are related to those of the input theory by
\be
\rho^{(p)}(T)=\rho(T^{\bar p}),\qquad \rho^{(p)}(S)=\rho(\sigma_p S).
\ee
These nice properties make it often possible to pair two Hecke images together to form a holomorphic CFT with $c=8k$, while the bilinear relation of the characters of the two  Hecke images goes to simple functions related to Klein $J$ function.

\subsection{Fermionic Hecke relations}
In this section, we define the fermionic Hecke operator for the NS characters of fermionic RCFTs. As we mentioned earlier, the reason why the NS sector is the most natural setting is that it transforms to itself under $S$ and $T^2$ of $\SL$, i.e., the NS characters form a $\Gamma_\theta$ vector-valued modular form of weight 0.  
To generalize from bosonic Hecke operator to the fermionic one, the key point is to define a \emph{fermionic transfer matrix} $\rho_{IJ}^{\rm F}( \sigma_p)$ as a sub-matrix of the bosonic one:
\be\label{submatrix}
\rho_{IJ}^{\rm F}(\sigma_p)\subset\rho_{ij}( \sigma_p).
\ee
Here $I,J$ are the indices of NS characters. This is required because fermionization and Hecke operation should be commutative. Most non-trivially, we find that the matrix $\rho_{IJ}^{\rm F}(\sigma_p)$ can be independently defined just by the NS data, i.e.,  $\rho^{\rm F}(S)$ and $\rho^{\rm F}(T^2)$, as
\be\label{eq:sigmaF}
\rho^{\rm F}(\sigma_p) =\rho^{\rm F}\Big( S^2(T^2)^{\frac{\bar{p}^2-\bar{p}}{2}}S (T^2)^{-\frac{{p}-1}{2}} S(T^2S)^{\bar{p}-1}\Big).
\ee
This combination is a slight modification of a function proposed very recently in \cite[Appendix B]{Cho:2022kzf} in the study of SMC.\footnote{In \cite{Cho:2022kzf}, the authors defined a function $H(p):=S^2(T^2)^{\frac{p^2-p}{2}}S (T^2)^{-\frac{\Bar{p}-1}{2}} S(T^2S)^{p-1}$. Three good properties were found there: $H(-1)=S^2$, $H(a)H(b)=H(ab)$ and $SH(a)=H(\bar{a})S$, for $a,b\in \IZ_N^\times$. This function is related to our $\sigma_p$ by $H(p)=\sigma_{\bar{p}}=\sigma_p^{-1}$.} 
It is well-defined owing to the fact that as long as $p$ is odd, $\frac{p^2-p}{2}$ and $\frac{\Bar{p}-1}{2}$ are always integers. A direct computation shows that
\be
\ba
S^2(T^2)^{\frac{\bar{p}^2-\bar{p}}{2}}S (T^2)^{-\frac{{p}-1}{2}} S(T^2S)^{\bar{p}-1}&=\left(
\begin{array}{cc}
\bar{p}(p \bar{p}-1)^2+\bar{p} & \ \ -(\bar{p}-1)^2 (p \bar{p}-1) \\
 p \bar{p}-1 & 1-p\bar{p}+p \\
\end{array}
\right)\\
&\equiv\left(
\begin{array}{cc}
\bar{p} & 0 \\
 0 & p \\
\end{array}
\right)\mod N.
\ea
\ee
Therefore this combination produces the correct $\sigma_p$ as the preimage $\mu_N^{-1}\mathrm{diag}(\bar{p},p)$. 
It should be emphasized that the conductor $N$ of a fermionic RCFT is always the same with the conductor of its bosonic theory. We can also define the conductor just by the NS data as the minimal positive integer $N$ such that $\rho^{\rm F}(T^2)^{N/2}=\rm Id$. Thus the conductor of any fermionic RCFT is always even. 
With these notions, we can introduce the following \emph{fermionic Hecke operator} $\mathsf{T}^{\rm F}_p$ for the NS characters $f^{\rm NS}_I$:
\be\label{Tp}
(\mathsf{T}^{\rm F}_p f^{\rm NS})_I(\tau) := \sum_J \rho_{IJ}^{\rm F}(\sigma_p) f^{\rm NS}_J(p \tau) +\sum_{b=0}^{p-1} f^{\rm NS}_I\Big(\frac{\tau+bN}{p}\Big) .
\ee
Apparently, this operator gives a similar map of the Fourier coefficients as  \eqref{Fouriermap}. Suppose 
$f_I(\tau)=\sum_n b_i(n)\,q^\frac{n}{N}$ and $(\TF_p f_I)(\tau)=\sum_{n} b^{(p)}_i(n)\,q^\frac{n}{N}$. Then we have
\begin{align}\label{FouriermapF}
\begin{split}
b^{(p)}_{I}(n) =
			\left\{
			\begin{array}{ll}
			p\, b_{I}(pn), &\ p \nmid n , \\
			p\, b_{I}(pn)+\sum_{J} \rho_{IJ}^{\rm F}(\sigma_p) b_J
			\big( \frac{n}{p}\big), &\ p \ | \ n.
			\end{array}
			\right.
		\end{split}
\end{align}
This operator can also be further generalized to non-prime $p$ with $\text{gcd}(p,N)=1$. Analogous to the bosonic case, we find
\begin{align}
\begin{split}
\left\{	\begin{array}{ll}
			\TF_{rs}=\TF_r\circ \TF_s,&\ \text{gcd}(r,s)=1,  \\
			\TF_{p^{n+1}}= \TF_{p}\circ\TF_{p^{n}}-p\,\sigma_p\circ\TF_{p^{n-1}},&\ p\text{ prime.}
			\end{array}
			\right.
		\end{split}
	\end{align}
These formulas hold owing to the important property found in \cite[Appendix B]{Cho:2022kzf} that for arbitrary $p,q$ coprime to $N$,
\be
\rho^{\rm F}(\sigma_p)\rho^{\rm F}(\sigma_q)=\rho^{\rm F}(\sigma_{pq}).
\ee
We call each $\rho^{\rm F}(\sigma_p)$ as a \emph{fermionic Hecke class}. Similar to the bosonic case, all fermionic Hecke classes form a finite abelian group. We will show such a finite abelian group for many examples later. 

We find that the fermionic Hecke operator has some basic properties resembling the bosonic Hecke operator:
\begin{itemize}
\item The \emph{multiple $p$ requirement}. The central charge $c^{(p)}$ and NS weights $ h^{(p)}_{\rm NS}$ of $\TF_p$ image of a theory with $(c,h_{\rm NS})$ satisfy the following property:
\be
c^{(p)}=p\, c,\quad\textrm{and}\quad  h^{(p)}_{\rm NS}\equiv p\, h_{\rm NS}\!\!\mod 1/2
\ee
By contrast, in the bosonic case, the conformal weights of the Hecke image satisfy multiple $p$ property mod $1$.
    \item The  \emph{homogeneous property}. We find that when $c^{\TF_p}\le 12$, any NS character of $\TF_p$ can be written as a degree $p$ homogeneous polynomial of the NS characters of the starting theory. When $c^{\TF_p}> 12$, this may not be always possible, but it is still always possible to write as the combination of a degree $p$ homogeneous polynomial and some lower degree homogeneous polynomials. This is slightly different from the bosonic cases, where the homogeneous property holds for arbitrary $c$.
\item The $S$ and $T^2$ matrices of the fermionic Hecke image $\TF_p$ are related to those of the input theory by
\be
\rho^{\textrm{F}}_{(p)}(T^2)=\rho^{\textrm{F}}(T^{2\bar p}),\qquad \rho^{\textrm{F}}_{(p)}(S)=\rho^{\textrm{F}}(\sigma_p S).
\ee
These properties make it sometimes possible to pair two fermionic Hecke images together to form a holomorphic SCFT with $c=12k$, while the bilinear relation of the characters of the two fermionic Hecke images gives a simple functions related to  $K(\tau)$ defined in \eqref{Ktau}.
\end{itemize}

To establish a Hecke relation, i.e., to identify a Hecke image with a known RCFT, it is always sufficient to check the Fourier coefficients of the characters up to a finite $q$ order. This is owing to the finite generation property of modular forms \cite{sturm}. In the fermionic cases, the same applies. For all fermionic Hecke relations we find in this paper, we have checked the Fourier coefficients of the NS characters to sufficiently high $q$ orders.

\label{sec:gHecke}
In our previous paper \cite[Section 2.5]{Duan:2022ltz}, we introduced the concept of \emph{generalized Hecke relations} $\T_p$ between bosonic theories for some $p$ not coprime to the conductor $N$. We find this concept is also useful in fermionic RCFTs. Analogously,  we define the \emph{generalized fermionic Hecke relations} $\TF_p$  by the following three conditions when $p$ is not coprime to the conductor $N$: 
\begin{enumerate}
   \item The central charge and NS conformal weights satisfy the multiple $p$ requirement.
\item The degeneracy for each non-vacuum NS primary is inherited.
\item The homogeneous property holds.
\end{enumerate}
The examples of generalized fermionic Hecke relations are not as rich as the bosonic cases. Nevertheless, we will encounter many such relations for free chiral fermions which will be discussed in the next subsection \ref{sec:1F} and for $SO(m)_1^3$ supersymmetric theories in Section \ref{sec:som13}. 
We observe from examples that generalized fermionic Hecke relations have similar properties as ordinary $\TF_p$, e.g., the Fourier coefficients of $\TF_p$ still satisfy the mod $p$ properties. However, the conductor $N$ will become $N/p$. These properties also exist for bosonic generalized Hecke relations \cite{Duan:2022ltz}.

\subsection{Example of free chiral ferimions}\label{sec:1F}
As a simple yet still nontrivial example, let us consider the fermionic Hecke images of a free Majorana fermion $F$. A free Majorana fermion is well-known to be the fermionization of 2d critical Ising model.  
The NS index $I$ only takes the vacuum 0 and the single NS character is just $\psi_{\rm NS}=\sqrt{\theta_3(\tau)/\eta(\tau)}$. The conductor $N=48$ is of course the same with the Ising model. The bosonic Hecke operation on the Ising model has been discussed in \cite{Harvey:2018rdc}, see also \cite[Table 9]{Duan:2022ltz}. Recall there are two Hecke classes forming $\IZ_2$ group for the bosonic Hecke operation of Ising model \cite{Harvey:2018rdc}. However, for NS character, the fermionic Hecke operation only has one single class, i.e., $\rho^{\rm F}(\sigma_p)=1$ for arbitrary admissible $p$. Besides, the homogeneous property implies the following simple relations
\be\label{TpF}
\T_{p}^{\rm F}=\begin{cases} (\T_{1}^{\rm F})^p,& p<24,  \\  
(\T_{1}^{\rm F})^p -p(\T_{1}^{\rm F})^{p-24}& 24<p<48 .
\end{cases}
\ee
This is just consistent with the fact that the fermionic Hecke image $\T_{p}^{\rm F}$ with $p<24$ describes the theory of $p$ free chiral fermions. Besides, the $\T^{\rm F}_p $ and $\T^{\rm F}_{24-p} $ fermionic Hecke images naturally form the $F_{24}$ SCFT of $c=12$. The bilinear relations of the NS characters of such pairs give the well-known identity $(\psi_{\rm NS})^{24} =K(\tau)+24$. We summarize the $c=12$ pairs in Table \ref{tb:HeckeIsing}. For $24<p<48$, the fermionic Hecke image $\T_{p}^{\rm F}$ no longer describes $p$ free fermions, but some more nontrivial interacting fermionic theories. For example, from the bosonic Hecke operation \cite{Harvey:2018rdc}, we can know that the $\T_{31}^{\rm F}$ and $\T_{35}^{\rm F}$ images should describe the fermionization of the WZW $(E_8)_2$ and $(C_{10})_1$ models. Moreover, the $\T_{47}^{\rm F}$ image should describe the 
SCFT associated with the Baby Monster group \cite{hoehn}. We have checked these are indeed correct.

\begin{table}[ht]
\def\arraystretch{1.1}
	\centering
	\begin{tabular}{|c|c|c|c|c|c|c|c|c|c|c|c|c|c|}
		\hline
		 $c$ & $h_{\rm NS}$  & $m_{1/2}$ & remark & $\tc$ & $\tilh_{\rm NS}$ &  $\tm_{1/2}$ & remark &$K(\tau)+n$ \\
		\hline
$\frac{p}{2}$ & $\frac{p}{16} $ & $p $ & $\T_{p}^{\rm F} $ & $\frac{24-p}{2}$ & $\frac{24-p}{16} $ & $ 24-p$ & $\T_{24-p}^{\rm F}$   &  $  24 $  \\
		\hline
		\end{tabular}
			\caption{(Generalized) fermionic Hecke images of one free chiral fermion.}
			\label{tb:HeckeIsing}
		\end{table}

In the above discussion, $p$ can only be those coprime to 48. It turns out we are not limited by such a condition. This is actually an excellent playground for generalized fermionic Hecke relations.  
Let us consider the theory of \emph{two} free fermions $2F$. This can be regarded as the fermionization of WZW $SO(2)_1$ model. The central charge doubles to $c=1$ and the conductor halves to $N=24$. The  single NS character is just $(\TF_1)^2$. Thus we can regard $2F$ as a generalized fermionic Hecke $\TF_2$ image of $F$. Consider the fermionic Hecke images of $2F$.  Obviously, there can only be one single class of $\rho^{\rm F}(\sigma_p)$, still as 1. Therefore, all relations in \eqref{TpF} and Table \ref{tb:HeckeIsing} still hold as long as for $p=2k,k\in\IZ$, we regard $\TF_p$ as generalized fermionic Hecke image $\T_k$ of $2F$.

As a step forward, let us consider the theory of \emph{three} free fermions $3F$. The central charge is $\frac32$. This theory can be regarded as the fermionization of WZW $SO(3)_1$ or equivalent $SU(2)_2$ model. The bosonic $SU(2)_2$ theory has conformal weights $0,\frac{3}{16},\frac12$. The relation between the NS character of $3F$ and $(A_1)_2$ affine characters is well-known to be 
\be
\chi_0^{3F}=(\TF_1)^3=\chi_0^{(A_1)_2}+\chi_{1/2}^{(A_1)_2}.
\ee
The conductor becomes $N=16$. The above relation also shows $\cF(A_1)_2$ can be regarded as a generalized fermionic Hecke $\TF_3$ image of $F$. We find equation \eqref{TpF} and Table \ref{tb:HeckeIsing} still hold if for $p=3k,k\in\IZ$, we regard $\TF_p$ as generalized fermionic Hecke image $\T_k$ of $3F$. For example, the $\TF_{21}F$ image, i.e., $\TF_7$ of $\cF(A_1)_2$ describes a theory $\cF(A_1)_2^7$ of central $\frac{21}{2}$. The single $\rm NS$ character of  $\cF(A_1)_2^7$ can be written as
\be
\chi_0^{\cF(A_1)_2^7}=\TF_7(\cF(A_1)_2)= (\TF_1)^{21}=(\chi_0^{(A_1)_2}+\chi_{1/2}^{(A_1)_2})^7   .
\ee
Together, $\cF(A_1)_2$ and $\cF(A_1)_2^7$ form the $c=12$ holomorphic SCFT $\cF(A_1)_2^8$ we mentioned in Section \ref{sec:coset}.

The theory of four free fermions $4F$ describes the fermionization of 
the double product $(A_1)_1^2$ owing to the appearance of weight $\frac12$ field. It is easy to check
\be
\chi_0^{4F}=(\TF_1)^4=(\chi^{(A_1)_1}_{0})^2+(\chi^{(A_1)_1}_{1/4})^2.
\ee
Thus, $\cF (A_1)_1^2$ can be regarded as a generalized $\TF_2$ image of $2F$ or a generalized $\T_4$ image of $F$. Moreover, the theories of $6F$, $8F$, $12F$ can be regarded as   generalized $\TF_6$, $\TF_8$ and $\TF_{12}$ images of $F$ which describe $ \cF (A_3)_1$, $\cF(D_4)_1$ and $\cF (D_6)_1$ respectively. In summary, \eqref{TpF} holds for all positive integers $p<48$, and Table \ref{tb:HeckeIsing} holds for all positive integers $p<24$.

As a final remark, we can also consider the $c=24$ pairs between $\TF_p$ and $\TF_{48-p}$. Suppose $1\le p\le 24$. The bilinear relation of $\TF_p$ and $\TF_{48-p}$ leads to a potential holomorphic SCFT of $c=24$ with single NS character 
\be
\TF_p\cdot \TF_{48-p}= (K(\tau)+24)^2-(48-p)(K(\tau)+24)=q^{-1}+pq^{-1/2}+24(p-1)+\dots.
\ee

\section{Two NS characters}\label{sec:2chi}
\subsection{Type $(SLY)_1$}\label{sec:SLY1}
Supersymmetric minimal model $SM(8,2)$ has 
$c=-\frac{21}{4}$ and $h=-\frac14$, while $SM_{\rm eff}(8,2)$ has $c_{\rm eff}=\frac{3}{4}$ and $h_{\rm eff}=\frac14$. It is well-known that $SM(8,2)$ can be realized as the fermionization of the bosonic minimal model $M(8,3)$. 
As $SM(8,2) $ is the simplest non-unitary supersymmetric minimal model, it was dubbed with the name ``supersymmetric Lee-Yang model". The effective theory by our notion is denoted as $(SLY)_1$.  It has the following two NS characters
\be
\ba
\chi_0&=q^{-\frac{1}{32}} \prod _{n=0}^{\infty} \frac{1}{\big(1-q^{\frac{1}{2} (8 n+1)}\big) \big(1-q^{\frac{1}{2} (8 n+4)}\big)\big(1-q^{\frac{1}{2} (8 n+7)}\big)},\\
 \chi_{\frac{1}{4}}&=q^{\frac{7}{32}} \prod _{n=0}^{\infty} \frac{1}{\big(1-q^{\frac{1}{2} (8 n+3)}\big) \big(1-q^{\frac{1}{2} (8 n+4)}\big) \big(1-q^{\frac{1}{2} (8 n+5)}\big)}.
\ea
\ee
Clearly the conductor $N=32$. The $S$-matrix and $T^2$-matrix are known to be 
\be\label{eq:SSLY1}
S=\left(
\begin{array}{cc}
 \cos \left(\frac{\pi }{8}\right) & \sin \left(\frac{\pi }{8}\right) \\
 \sin \left(\frac{\pi }{8}\right) & -\cos \left(\frac{\pi }{8}\right) \\
\end{array}
\right), \quad T^2=\left(
\begin{array}{cc}
 e^{-\frac{\pi i}{8}} & 0 \\
 0 & e^{\frac{7\pi i}{8}} \\
\end{array}
\right).
\ee  
This $S$ matrix leads to the NS fusion rule $\phi_1\times \phi_1=\phi_0-2\phi_1$. We notice the two NS characters satisfy the following degree 8 homogeneous polynomial identity
\be\label{SLY1id}
\chi_0\chi_{\frac{1}{4}}\big(\chi_0^6-7\chi_0^4\chi_{\frac{1}{4}}^2+7\chi_0^2\chi_{\frac{1}{4}}^4-\chi_{\frac{1}{4}}^6\big)=1.
\ee
This resembles the famous Ramanujan identity of the two Lee-Yang characters, see e.g. \cite[Equation (2.48)]{Duan:2022ltz}. The two $\rm R$ characters can be found in e.g. \cite{Bae:2020xzl}. The $\rm \widetilde{R}$ character is just a constant 1.

Let us consider the fermionic Hecke operation of $(SLY)_1$. 
We find there are in total four classes for $\rh$ for $p$ modulo 16 which are represented by
\be\nonumber
\rho^{\rm F}(\sigma_{1,15})=\!\left(
\begin{array}{cc}
 1 & 0 \\
 0 & 1 \\
\end{array}
\right)\!,\ 
\rho^{\rm F}(\sigma_{3,13})=\!\left(
\begin{array}{cc}
 0 & -1 \\
 1 & 0 \\
\end{array}
\right)\!,\ 
\rho^{\rm F}(\sigma_{5,11})=\!\left(
\begin{array}{cc}
 0 & 1 \\
- 1 & 0 \\
\end{array}
\right)\!,\ 
\rho^{\rm F}(\sigma_{7,9})=\!\left(
\begin{array}{cc}
 -1 & 0 \\
 0 & -1 \\
\end{array}
\right)\!.
\ee
Clearly, the four fermionic Hecke classes form a $\IZ_4$ group, which is just the $\IZ_4$ group of $p$ with the same quadratic residue modulo 32. With the above $\rho^{\rm F}(\sigma_{p})$, 
we compute all fermionic Hecke images $\TF_p$ up to $p<16$ and organize them in pairs w.r.t $c=12$ in Table \ref{tb:HeckeSLY1}. These fermionic images are consistent with the computation on bosonic Hecke images of $M(8,3)$ in \cite[Table 38]{Duan:2022ltz}. 
\begin{table}[ht]
\def\arraystretch{1.1}
	\centering
	\begin{tabular}{|c|c|c|c|c|c|c|c|c|c|c|c|c|c|}
		\hline
		 $c$ & $h_{\rm NS}$  & $m_{1/2}$ & remark & $\tc$ & $\tilh_{\rm NS}$ &  $\tm_{1/2}$ & remark & $K(\tau)+n$ \\
		\hline
$\frac{3}{4}$ & $\frac{1}{4} $ & $1 $ & $\T_{1}^{\rm F},(SLY)_1 $ & $\frac{45}{4}$ & $\frac34 $ & $15 $ & $\T_{15}^{\rm F}$   &  $  16 $  \\
$\frac{9}{4}$ & $\frac{1}{4} $ & $0 $ & $\T_{3}^{\rm F},\cF(A_{1})_6 $ & $\frac{39}{4}$ & $\frac{3}{4} $ & $ 0$ & $\T_{13 }^{\rm F},\cF(C_{6})_1$   &  $  0 $  \\
$\frac{ 15}{4}$ & $\frac{ 1}{4} $ & $5 $ & $\T_{5}^{\rm F},\star $ & $\frac{33 }{4}$ & $\frac{3 }{4} $ & $11  $ & $\T_{11 }^{\rm F},\star$   &  $  -16 $  \\
$\frac{21 }{4}$ & $\frac{1 }{4} $ & $ 14$ & $\T_{7}^{\rm F},\star $ & $\frac{27 }{4}$ & $\frac{ 3}{4} $ & $ 18 $ & $\T_{9 }^{\rm F},\star$   &  $  -32 $  \\
		\hline
	
		\end{tabular}
			\caption{Fermionic Hecke images of supersymmetric Lee-Yang model $(SLY)_1$. }
			\label{tb:HeckeSLY1}
		\end{table}
		
For example, we find that the fermionic Hecke image $\T^{\rm F}_3$ of $(SLY)_1$ describes exactly the fermionization of $(A_1)_6$. This is a renowned RCFT with emergent supersymmetry with true supersymmetric vacua. The two NS characters are related to the affine characters and the $(SLY)_1$ characters by the following combination:
\be
\ba
f^{\cF(A_1)_6}_0 & = \chi^{(A_1)_6}_0 + \chi^{(A_1)_6}_{\frac32} =\chi_0^3-3\chi_0\chi_{\frac14}^2   ,  \\ 
  f^{\cF(A_1)_6}_{\frac14} & = \chi^{(A_1)_6}_{\frac14} + \chi^{(A_1)_6}_{\frac34} 
  = 3\chi_0^2\chi_{\frac14}-\chi_{\frac14}^3. 
  \ea
  \ee
Besides, we find the fermionic Hecke image $\T^{\rm F}_{13}$ describes the fermionization of $(C_6)_1$. To be precise, we find the following relations
\be\nonumber
\ba
f^{\cF(C_6)_1}_0 & = \chi^{(C_6)_1}_0 + \chi^{(C_6)_1}_{\frac32} =  \chi_0^3 (\chi_0^{10}-13 \chi_0^8 \chi_{\frac14}^2+130 \chi_0^6 \chi_{\frac14}^4-338 \chi_0^4 \chi_{\frac14}^6+221 \chi_0^2 \chi_{\frac14}^8-65 \chi_{\frac14}^{10})  ,  \\
  f^{\cF(C_6)_1}_{\frac34} & = \chi^{(C_6)_1}_{\frac34} + \chi^{(C_6)_1}_{\frac54} 
  = \chi_{\frac14}^3 (65 \chi_0^{10}-221 \chi_0^8 \chi_{\frac14}^2+338 \chi_0^6 \chi_{\frac14}^4-130 \chi_0^4 \chi_{\frac14}^6+13 \chi_0^2 \chi_{\frac14}^8-\chi_{\frac14}^{10}) . 
  \ea
  \ee
The $\cF (C_6)_1$ is also a famous example with emergent supersymmetry with supersymmetric vacua, see e.g. \cite{Johnson-Freyd:2019wgb} and the fermionic characters in \cite[Table 14]{Bae:2021lvk}. 
As a $c=12$ pair, these two fermionic Hecke images form the Conway SCFT. Their NS characters give the following bilinear identity
\be\label{KSLY1}
\ba
K(\tau)=\TF_3\cdot\TF_{13}=&\,\chi_{0}^{16}-16 \chi_{0}^{14} \chi_{\frac14}^2+364 \chi_{0}^{12} \chi_{\frac14}^4-1456 \chi_{0}^{10} \chi_{\frac14}^6+2470 \chi_{0}^8 \chi_{\frac14}^8\\
&-1456 \chi_{0}^6 \chi_{\frac14}^{10}+364 \chi_{0}^4 \chi_{\frac14}^{12}-16 \chi_{0}^2 \chi_{\frac14}^{14}+\chi_{\frac14}^{16}.
\ea
\ee 

The $\TF_{15}$ image is not unitary, but it may serve as a certain fermionic analogy of the famous WZW $(E_{7\frac12})_1$ theory. Therefore we show its character in the following:
\be
\ba
f^{\TF_{15}}_0 & = \chi_{0}^3(\chi_{0}^{12}+105 \chi_{0}^8 \chi_{\frac14}^4-280 \chi_{0}^6 \chi_{\frac14}^6+435 \chi_{0}^4 \chi_{\frac14}^8-168 \chi_{0}^2 \chi_{\frac14}^{10}+35 \chi_{\frac14}^{12})\\
&=q^{-\frac{15}{32}}(1+15 \sqrt{q}+225 q+1555 q^{3/2}+7920 q^2+32580 q^{5/2}+\dots),  \\
  f^{\TF_{15}}_{\frac34} & = \chi_{\frac14}^3 (35 \chi_{0}^{12}-168 \chi_{0}^{10} \chi_{\frac14}^2+435 \chi_{0}^8 \chi_{\frac14}^4-280 \chi_{0}^6 \chi_{\frac14}^6+105 \chi_{0}^4 \chi_{\frac14}^8+\chi_{\frac14}^{12})\\
&=q^{\frac{9}{32}}(35+252 \sqrt{q}+1485 q+6805 q^{3/2}+25845 q^2+86220 q^{5/2}+\dots) . 
  \ea
  \ee
The $\TF_{15}$ and $(SLY)_1$ do not pair as a consistent holomorphic SCFT of $c=12$. However, from $K(\tau)+16=\TF_1\cdot \TF_{15}$ we can obtain a different expression for $K(\tau)$. The difference from \eqref{KSLY1} is exactly 16 times the square of identity \eqref{SLY1id}.

\subsection{Type $SM(5,3)$}\label{sec:SM53}
$SM(5,3)$ is the simplest unitary supersymmetric minimal model.  It can be realized as the fermionization of the bosonic minimal model $M(5,4)$ \cite{Friedan:1983xq,Friedan:1984rv,Bershadsky:1985dq}. It has central charge $c=\frac{7}{10}$ and two NS primaries with weights $0$ and $\frac{1}{10}$. The conductor $N=240$. The $S$-matrix for the two NS characters is 
\be\label{SSM53}
S=\frac{1}{\sqrt{30}}\left(
\begin{array}{cc}
\alpha_- & \alpha_+ \\
\alpha_+ & -\alpha_- \\
\end{array}
\right) ,\qquad \alpha_\pm=\sqrt{15\pm 3\sqrt{5}}.  
    \ee
    This leads to the simple NS fusion rule $\phi_1\times \phi_1=\phi_0+\phi_1$.

In our previous paper \cite{Duan:2022ltz} we have computed the bosonic Hecke images of $M(5,4)$ theory, see Table 31 therein. 
Now let us purely consider the fermionic Hecke operations. Contrast to the eight bosonic Hecke classes, we find there are in total four classes for the fermionic Hecke operation of  $SM(5,3)$: for $p\equiv 1, 19, 41, 59, 61, 79, 101, 119\!\!\mod 120$, $\rho(\sigma_{p})^{\rm F}=\mathrm{Id}$, for $p\equiv 11, 29, 31, 49, 71, 89, 91, 109 \!\!\mod 120$, $\rho(\sigma_{p})^{\rm F}=-\mathrm{Id}$,
    \be
    \ba
&\textrm{for }  p\equiv7, 13, 47, 53, 67, 73, 107, 113\!\!\mod 120,\quad 
\rho(\sigma_{p})^{\rm F}=\left(
\begin{array}{cc}
 0 & 1 \\
 -1 & 0 \\
\end{array}
\right),\\ 
&\textrm{for }  p\equiv 17, 23, 37, 43, 77, 83, 97, 103 \!\!\mod 120,\quad 
\rho(\sigma_{p})^{\rm F}=\left(
\begin{array}{cc}
 0 & -1 \\
 1 & 0 \\
\end{array}
\right).
\ea
\ee
The four fermionic Hecke classes form a $\IZ_4$ group. We remark that this $\IZ_4$ is actually different from the $\IZ_4$ group of $p$ with the same quadratic residue modulo 240. For example, it is easy to check that all $p$ satisfying $p^2\equiv 1\ \mathrm{mod}\ 240$ are $p=1, 31, 41, 49, 71, 79, 89, 119\ \mathrm{mod}\ 120$. 
We compute all fermionic Hecke images $\TF_p$ for $p<24$ and organize them in Table \ref{tb:SM53}. 
These are all consistent with the computation on the bosonic Hecke images of $M(5,4)$ in \cite{Duan:2022ltz}. Notably, the $\TF_{19}$ image describes the fermionization of WZW $(E_7)_2$ theory. The $\cF (E_7)_2$ is a well-known example with emergent supersymmetry, yet having a non-supersymmetric Ramond ground state.  The relation between the NS characters of $\cF (E_7)_2$ and affine $(E_7)_2$ characters can be found in e.g. \cite[Equation (5.94)]{Bae:2020xzl}.
\begin{table}[ht]
\def\arraystretch{1.1}
	\centering
	\begin{tabular}{|c|c|c|c|c|c|c|c|c|c|c|c|c|c|}
		\hline
		 $c$ & $h_{\rm NS}$  & $m_{1/2}$ & remark   \\
		\hline
$\frac{7}{10}$ & $\frac{ 1}{10} $ & $ 0$ & $\T_{1}^{\rm F}, SM(5,3) $   \\
$\frac{49}{10}$ & $\frac{ 1}{5} $ & $ 7 $ & $\T_{7}^{\rm F},\star $   \\
$\frac{77}{10}$ & $\frac{ 3}{5} $ & $ 11 $ & $\T_{11}^{\rm F},\star $   \\
$\frac{91}{10}$ & $\frac{4 }{5} $ & $ 13 $ & $\T_{13}^{\rm F},\star $   \\
$\frac{119}{10}$ & $\frac{ 7}{10} $ & $ 17 $ & $\T_{17}^{\rm F},\star $   \\
$\frac{133}{10}$ & $\frac{9 }{10} $ & $ 0 $ & $\T_{19}^{\rm F} ,\cF(E_7)_2 $   \\
$\frac{161}{10}$ & $\frac{ 4}{5} $ & $  0$ & $\T_{23}^{\rm F} $   \\
		\hline
	
		\end{tabular}
			\caption{Fermionic Hecke images of $SM(5,3)$.}
			\label{tb:SM53}
		\end{table}
		
It is straightforward to further compute the fermionic Hecke images of double product $SM(5,3)^2$, which produce lots of three NS-character theories with degeneracy $(1,2,1)$.  The conductor halves to $N=120$. As a remark, we notice that $SM(5,3)^2$ describes exactly the fermionization of $\mathbb{Z}_8$ parafermion CFT discussed in \cite[Equation (A.3)]{Bae:2021lvk}. 

\subsection{Type $SM_{\rm sub}(60,2)$}
We now consider a theory rather similar to $SM(5,3)$, but non-unitary. 
The supersymmetric minimal model $SM(60,2)$ has 
$c=-\frac{413}{5}$ and $c_{\rm eff}=\frac{7}{5}$. The effective theory, i.e., $(SLY)_{14}$ has 15 NS characters $\chi_i,i=1,2,\dots,14$ with weights
\be
0,\frac{1}{30},\frac{1}{10},\frac{1}{5},\frac{1}{3},\frac{1}{2},\frac{7}{10},\frac{14}{15},\frac{6}{5},\frac{3}{2},\frac{11}{6},\frac{11}{5},\frac{13}{5},\frac{91}{30},\frac{7}{2}.
\ee
We construct a non-diagonal $\Gamma_\theta$ modular invariant of $E_8$ type out of $(SLY)_{14}$ by
\be
\ba
\chi_0^{SM_{\rm sub}(60,2)}&=\chi_0+ \chi_{5}+ \chi_{9}+ \chi_{14}=q^{-\frac{7}{120}}(1+2 \sqrt{q}+2 q+4 q^{3/2}+6 q^2+8 q^{5/2}+\dots),\\ \chi_{1/5}^{SM_{\rm sub}(60,2)}&=\chi_3+ \chi_{6}+ \chi_{8}+ \chi_{11}=q^{\frac{17}{120}}(1+2 \sqrt{q}+q+2 q^{3/2}+5 q^2+6 q^{5/2}+\dots).
\ea
\ee
Note the 2 times of subscripts $i$ of $\chi_i$ plus 1 are exactly the eight exponents of $E_8$ Lie algebra. It is easy to check 
\be
Z_{\rm NS}^{SM_{\rm sub}(60,2)}=\big|\chi_0^{SM_{\rm sub}(60,2)}\big|^2+\big|\chi_{1/5}^{SM_{\rm sub}(60,2)}\big|^2
\ee
is $\Gamma_\theta$ modular invariant. The $S$-matrix can be easily deduced from the one of $(SLY)_{14}$ as
\be\label{SSM602}
S=\frac{1}{\sqrt{30}}\left(
\begin{array}{cc}
\alpha_+ & \alpha_- \\
\alpha_- & -\alpha_+ \\
\end{array}
\right) ,\qquad \alpha_\pm=\sqrt{15\pm 3\sqrt{5}}.
\ee
Note $S^2=\rm Id$.  We denote this sub-theory of $(SLY)_{14}$ as $SM_{\rm sub}(60,2)$. Clearly the conductor $N=120$.  Notice that the $S$-matrix here is just different from the one \eqref{SSM53} of $SM(5,3)$ by exchanging the plus and minus in the subcripts of $\alpha$. This leads to the non-unitary NS fusion rule $\phi_1\times \phi_1=\phi_0-\phi_1$.

Consider the fermionic Hecke images of this $E_8$ type sub-theory.  We find there are in total four classes for the fermionic Hecke operation of the NS characters: for $p\equiv 1, 19, 41, 59\!\!\mod 60$, $\rho^{\rm F}(\sigma_{p})=\mathrm{Id}$, for $p\equiv11, 29, 31, 49  \!\!\mod 60$, $\rho^{\rm F}(\sigma_{p})=-\mathrm{Id}$,
    \be
    \ba
&\textrm{for }  p\equiv17, 23, 37, 43\!\!\mod 60,\quad 
\rho^{\rm F}(\sigma_{p})=\left(
\begin{array}{cc}
 0 & 1 \\
 -1 & 0 \\
\end{array}
\right),\\ 
&\textrm{for }  p\equiv 7, 13, 47, 53\!\!\mod 60,\quad 
\rho^{\rm F}(\sigma_{p})=\left(
\begin{array}{cc}
 0 & -1 \\
 1 & 0 \\
\end{array}
\right).
\ea
\ee
The four fermionic Hecke classes still form a $\IZ_4$ group. 
We remark that the first two classes of $p$ have quadratic residue $1$ modulo 120, while the last two classes have quadratic residue $49$ modulo 120. We compute all fermionic Hecke images $\TF_p$ up to $p<20$ and organize them in Table \ref{tb:SM602}. Interestingly, we find the $\TF_{13}$ image describes exactly the fermionization of WZW $(D_7)_3$. The relation between the NS characters of $\cF(D_7)_3$ and the affine characters can be found in e.g. \cite[Equation (5.99)]{Bae:2020xzl}. We remark that $\cF(D_7)_3$ and $SM(5,3)$ belong to the same object in the super modular category of rank 4.
\begin{table}[ht]
\def\arraystretch{1.1}
	\centering
	\begin{tabular}{|c|c|c|c|c|c|c|c|c|c|c|c|c|c|}
		\hline
		 $c$ & $h_{\rm NS}$  & $m_{1/2}$ & remark   \\
		\hline
$\frac{7}{5}$ & $\frac{ 1}{5} $ & $ 2 $ & $\T_{1}^{\rm F} ,SM_{\rm sub}(60,2)$   \\
$\frac{49}{5}$ & $\frac{ 2}{5} $ & $ 14 $ & $\T_{7}^{\rm F} $   \\
$\frac{77}{5}$ & $\frac{ 7}{10} $ & $ 0 $ & $\T_{11}^{\rm F},\star $   \\
$\frac{91}{5}$ & $\frac{ 11}{10} $ & $ 0 $ & $\T_{13}^{\rm F},\cF(D_7)_3 $   \\
$\frac{119}{5}$ & $\frac{7 }{5} $ & $ 0 $ & $\T_{17}^{\rm F},\star $   \\
$\frac{133}{5}$ & $\frac{ 13}{10} $ & $  0$ & $\T_{19}^{\rm F} $   \\
		\hline
	
		\end{tabular}
			\caption{Fermionic Hecke images of $SM_{\rm sub}(60,2)$.}
			\label{tb:SM602}
		\end{table}

\subsection{Type $\cS A_1$}
The $\cN=1$ supersymmetric $A_1$ is the supersymmetrization of the lattice $A_1$ CFT by coupling it with one free chiral fermion. Clearly it has central charge $c=\frac32$ and NS weights $0,\frac14$. The fermionic Hecke operation for this case is almost trivial as it is  parallel to the bosonic Hecke operation for WZW $(A_1)_1$ theory. However, it is interesting to observe how much the supersymmetrization changes the theory. For example, the conductor $N$ is changed from 24 to 16. Nevertheless, the $S$-matrix remains the same. Consider the fermionic Hecke operation $\TF_p$ for  $\cS A_1$. We find there exist in total two classes $\rho^{\rm F}(\sigma_p)$ for $p\equiv1,7\!\!\mod 8$, $ \rh=\rm Id$, while for $p\equiv3,5\!\!\mod 8$, $\rh=-\rm Id$. We compute all fermionic Hecke images $\TF_p$ up to $p<8$ and organize them in pairs w.r.t $c=12$ in Table \ref{tb:HeckeSA1}.  Just like the bosonic Hecke operation $\T_7(A_1)_1= (E_7)_1$ \cite{Harvey:2018rdc}, the fermionic Hecke operation produces $\T_7\cS A_1= \cS  E_7$. Together $\cS A_1$ and  $\cS  E_7$ form a holomorphic SCFT of $c=12$ that is $\cS E_8$. On the other hand,  $\T_{3}^{\rm F}$ and $\T_{5}^{\rm F}$ images have negative NS vacuum, thus are unphysical theories.
\begin{table}[ht]
\def\arraystretch{1.1}
	\centering
	\begin{tabular}{|c|c|c|c|c|c|c|c|c|c|c|c|c|c|}
		\hline
	 $c$ & $h_{\rm NS}$  & $m_{1/2}$ & remark & $\tc$ & $\tilh_{\rm NS}$ &  $\tm_{1/2}$ & remark & $K(\tau)+n$ \\
		\hline
$\frac{ 3}{2}$ & $\frac{1 }{4} $ & $1 $ & $\T_{1}^{\rm F} ,\cS A_1$ & $\frac{21 }{2}$ & $ \frac34$ & $ 7 $ & $\T_{7 }^{\rm F},\cS E_7$   &  $ 8  $  \\
$\frac{ 9}{2}$ & $\frac{ 1}{4} $ & $ 9$ & $\T_{3}^{\rm F},\star $ & $\frac{ 15}{2}$ & $\frac{ 3}{4} $ & $  15$ & $\T_{5 }^{\rm F},\star$   &  $ -24  $  \\
		\hline
	
		\end{tabular}
			\caption{Fermionic Hecke images of supersymmetric $A_1$ theory $\cS A_1$. }
			\label{tb:HeckeSA1}
		\end{table}

\subsection{Type $\cS^2 A_1$}\label{sec:S2A1}
The $\cN=2$ supersymmetric $A_1$ theory $\cS^2 A_1$ is the first one of a unitary $\cN=2$ series. It has central charge $c=1$ and NS weights $0,(\frac16)_2$. Note the non-vacuum NS primary has degeneracy two. This theory can be realized as a non-diagonal modular invariant of $SM(6,4)$ minimal model, see e.g. \cite[Equation (5.29)]{Bae:2020xzl} for the NS characters. It can also be regarded as the fermionization of bosonic $U(1)_6$ theory. The conductor $N=24$. The full $S$-matrix for the three NS primaries is well-known to be the same with the $S$-matrix of WZW $(A_2)_1$ as
\be
S=\frac{1}{\sqrt{3}}\left(
\begin{array}{ccc}
 1 & 1 & 1 \\
 1 & \omega_1 & \omega_1^2 \\
 1 & \omega_1^2 & \omega_1 \\
\end{array}
\right),\qquad \omega_1=e^{\frac{2 i \pi }{3}}  .
\ee
It is also equivalent to consider a reduced $S$-matrix for just two NS characters:
\be\label{eq:SreducedN2A1}
S_{\rm reduced}=\frac{1}{\sqrt{3}}\left(
\begin{array}{cc}
 1 & 2\\
 1 & -1 \\
\end{array}
\right).
\ee

Let us study the fermionic Hecke operation for the reduced $S$-matrix as it is computationally easier. We find there exist  two classes of  $\rho^{\rm F}(\sigma_p)$. For $p\equiv1,11\!\!\mod 12$, $ \rh=\rm Id$, while for $p\equiv5,7\!\!\mod 12$, $ \rh=-\rm Id$. We compute all fermionic Hecke images $\TF_p$ for $p<12$ and organize them in pairs w.r.t $c=12$ in Table \ref{tb:N2A1}. 
These are consistent with the computation on bosonic Hecke images of $U(1)_6$ theory in our previous work \cite[Table 37]{Duan:2022ltz}. Notably, the $\TF_{11}$ image describes the fermionization of WZW $(A_{11})_1$ model. The relation between the NS characters of $\cF (A_{11})_1$ and affine characters can be found in e.g. \cite[Equation (3.12)]{Bae:2021lvk}. The $\cF (A_{11})_1$ theory has unbroken supersymmetry, and its automorphism group is known to be related to $Suz:2$ sporadic group \cite{Johnson-Freyd:2019wgb}. The homogeneous property of fermionic Hecke relation implies the following nice relations between the NS characters of $\cS^2 A_1$ and $\cF(A_{11})_1$:
\be
\ba
\chi_0^{\cF(A_{11})_1}&=\chi_0^{11}+132\chi_0^{5}\chi_{1/6}^6 +110\chi_0^{2}\chi_{1/6}^9,\\
\chi_{5/6}^{\cF(A_{11})_1}&=3(4\chi_{5/6}^{11}+22\chi_0^{6}\chi_{1/6}^5 +55\chi_0^{3}\chi_{1/6}^8).
\ea
\ee
Here we use $\chi_0$ and $\chi_{1/6}$ to denote the two NS characters of $\cS^2A_1$. Together they form the bilinear identity 
\be
\ba
K(\tau)&=\TF_1\cdot \TF_{11}=\chi_0\chi_0^{\cF(A_{11})_1}+2\chi_{1/6}\chi_{5/6}^{\cF(A_{11})_1}\\
&=\chi_0^{12}+264\chi_0^{6}\chi_{1/6}^6 +440\chi_0^{3}\chi_{1/6}^9+24\chi_{1/6}^{12}.
\ea
\ee
Different $c=12$ pairs lead to a simple identity 
\be
\chi_{1/6}(\chi_0^3-\chi_{1/6}^3)=1.
\ee
It is easy to prove this identity from the $S$-matrix \eqref{eq:SreducedN2A1}.
\begin{table}[ht]
\def\arraystretch{1.1}
	\centering
	\begin{tabular}{|c|c|c|c|c|c|c|c|c|c|c|c|c|c|}
		\hline
		 $c$ & $h_{\rm NS}$  & $m_{1/2}$ & remark & $\tc$ & $\tilh_{\rm NS}$ &  $\tm_{1/2}$ & remark & $K(\tau)+n$ \\
		\hline
$1$ & $\frac{ 1}{6} $ & $ 0$ & $\T_{1}^{\rm F}, \cS^2 A_1 $ & $11$ & $\frac{5 }{6} $ & $  0$ & $\T_{ 11}^{\rm F},\cF(A_{11})_1$   &  $  0 $  \\
$5$ & $\frac{1 }{3} $ & $ 10$ & $\T_{5}^{\rm F},\star $ & $7$ & $\frac{ 2}{3} $ & $ 14 $ & $\T_{ 7}^{\rm F},\star$   &  $  -24 $  \\
		\hline
	
		\end{tabular}
			\caption{Fermionic Hecke images of $\cS^2 A_1$. The non-vacuum $\rm NS$ character has degeneracy 2.}
			\label{tb:N2A1}
		\end{table}

Consider a product type theory composed of a $\cS^2 A_1$ and a free chiral fermion. The central charge $c=\frac32$. Adding a free fermion does not change the $S$-matrix of $\cS^2 A_1$, but changes its $T^2$-matrix and conductor. The two NS characters have the following Fourier expansion
\be
\ba
\chi_0&= q^{-\frac{1}{16}}(1+\sqrt{q}+q+4 q^{3/2}+5 q^2+6 q^{5/2}+9 q^3 +\dots) ,\\
\chi_{1/6}&=q^\frac{5}{48}( 1+2 \sqrt{q}+2 q+3 q^{3/2}+5 q^2+8 q^{5/2}+10 q^3+\dots).
\ea
\ee
The second character has degeneracy two. The conductor becomes $N=48$.

Let us study the fermionic Hecke operation for this new theory. We find there exist  two classes $\rho^{\rm F}(\sigma_p)$ for $p\equiv 1, 11, 13, 23\!\!\mod 24$, $ \rh=\rm Id$, while for $p\equiv 5, 7, 17, 19\!\!\mod 24$, $ \rh=-\rm Id$. We compute all fermionic Hecke images $\TF_p$ up to $p< 14$ and summarize them in Table \ref{tb:N2A1F}. 
Interestingly, we notice the $\TF_{13}$ image describes exactly the fermionization of WZW $(E_6)_4$ model. The relation between the NS characters of $\cF(E_6)_4$ and the affine characters can be found in for example \cite[Equation (5.104)]{Bae:2020xzl}. 

\begin{table}[ht]
\def\arraystretch{1.1}
	\centering
	\begin{tabular}{|c|c|c|c|c|c|c|c|c|c|c|c|c|c|}
		\hline
		 $c$ & $h_{\rm NS}$  & $m_{1/2}$ & remark   \\
		\hline
$\frac32$ & $\frac{ 1}{6} $ & $ 0$ & $\T_{1}^{\rm F}, \cS^2 A_1\otimes F $   \\
$\frac{15}{2}$ & $\frac{1 }{3} $ & $ 5 $ & $\T_{5}^{\rm F},\star $  \\
$\frac{21}{2}$ & $\frac{2 }{3} $ & $ 7 $ & $\T_{7}^{\rm F},\star $  \\
$\frac{33}{2}$ & $\frac{ 5}{6} $ & $ 0 $ & $\T_{11}^{\rm F}$  \\
$\frac{39}{2}$ & $\frac{ 7}{6} $ & $ 0 $ & $\T_{13}^{\rm F},\cF(E_6)_4 $  \\
		\hline
	
		\end{tabular}
			\caption{Fermionic Hecke images of $\cS^2 A_1\otimes F$. The non-vacuum $\rm NS$ character has degeneracy 2.}
			\label{tb:N2A1F}
		\end{table}

\section{Three NS characters}\label{sec:3chi}

\subsection{Type $(SLY)_2$}
Supersymmetric minimal model $SM(12,2)$ has $c=-11$ and $h=-\frac13,-\frac12$, while the effective theory $SM_{\rm eff}(12,2)$, i.e., $(SLY)_2$ has $c_{\rm eff}=1$ and $h_{\rm eff}=\frac16,\frac12$. The three NS characters of $(SLY)_2$ can be easily computed from \eqref{chisum} or \eqref{chiprod}. The conductor $N=24$. We find 
the $S$-matrix of the three NS characters to be
\be
S=\frac{1}{2\sqrt{3}}\left(
\begin{array}{ccc}
 \sqrt{3}+1 & 2 & \sqrt{3}-1 \\
 2 & -2 & -2 \\
 \sqrt{3}-1 & -2 & \sqrt{3}+1 \\
\end{array}
\right) .
\ee
It is easy to check $S^2=\rm Id$. This $S$ matrix gives the NS fusion rules $\phi_1\times \phi_1=\phi_0-\phi_1+\phi_2$, $\phi_1\times \phi_2=\phi_1-2\phi_2$ and $\phi_2\times \phi_2=\phi_0-2\phi_1+2\phi_2$.

Consider the fermionic Hecke operation on $(SLY)_2$. We find there exist  two classes $\rho^{\rm F}(\sigma_p)$ for $p\!\!\mod 12$, 
\be
\rho^{\rm F}(\sigma_{1,11})={\rm Id},\ \textrm{   and   }\ \rho^{\rm F}(\sigma_{5,7})=\left(
\begin{array}{ccc}
 0 & 0 & 1 \\
 0 & -1 & 0 \\
 1 & 0 & 0 \\
\end{array}
\right).
\ee
Obviously, they form a $\IZ_2$ group. 
We compute all fermionic Hecke images $\TF_p$ for $p< 24$ and summarize those with $p<12$ in $c=12$ pairs in Table \ref{tb:SLY2}. Notably, we find the $\TF_5$ image describes exactly the fermionization of WZW $(B_2)_3$ model, while the $\TF_7$ image describes exactly the fermionization of WZW $(C_3)_2$ model. Together they form the character of holomorphic Conway SCFT of $c=12$. For the explicit relations between the NS characters of $\cF (B_2)_3$ and affine characters, we refer to \cite[Equation (3.71)]{Bae:2021mej}, while for $\cF (C_3)_2$ we refer to \cite[Equation (3.73)]{Bae:2021mej}. From the homogeneous property of fermionic Hecke images, we find that the NS character of $\cF (B_2)_3$ can be written as
\be
\ba
\chi^{\cF (B_2)_3}_0&=\chi_0(\chi_0^4-5 \chi_{1/2} \chi_0^3+15 \chi_{1/2}^2 \chi_0^2-5 \chi_{1/2}^3 \chi_0+10 \chi_{1/2}^4 ),\\ 
\chi^{\cF (B_2)_3}_{1/3}&=
\chi_{1/6}^2 (5 \chi_{0}^3-15 \chi_{1/2} \chi_{0}^2+15 \chi_{1/2}^2 \chi_{0}+4 \chi_{1/6}^3-5 \chi_{1/2}^3)
 \\
\chi^{\cF (B_2)_3}_{1/2}&= \chi_{1/2} (10 \chi_{0}^4-5 \chi_{1/2} \chi_{0}^3+15 \chi_{1/2}^2 \chi_{0}^2-5 \chi_{1/2}^3 \chi_{0}+\chi_{1/2}^4).
\ea
\ee
However, unlike the $(SLY)_1$ case where the homogeneous polynomial of a fermionic Hecke image is unique, here the expression is no longer unique due to vanishing identities like
\be
0=\chi_{1/2} \chi_0^3+\chi_{1/2}^3 \chi_0-\chi_0 \chi_{1/6}^3+\chi_{1/6}^3 \chi_{1/2}.
 \ee
This kind of vanishing identities of characters also begin to appear in the bosonic minimal model $(LY)_2$, i.e. $M_{\rm eff}(7,2)$.

\begin{table}[ht]
\def\arraystretch{1.1}
	\centering
	\begin{tabular}{|c|c|c|c|c|c|c|c|c|c|c|c|c|c|}
		\hline
		 $c$ & $h_{\rm NS}$  & $m_{1/2}$ & remark & $\tc$ & $\tilh_{\rm NS}$ &  $\tm_{1/2}$ & remark & $K(\tau)+n$ \\
		\hline
$1$ & $\frac{ 1}{6},\frac12 $ & $ 1$ & $\T_{1}^{\rm F}, (SLY)_2 $ & $11$ & $\frac12,\frac{5 }{6} $ & $  11$ & $\T_{ 11}^{\rm F} $   &  $  12 $  \\
$5$ & $\frac{1 }{3},\frac12 $ & $ 0$ & $\T_{5}^{\rm F},\cF(B_2)_3 $ & $7$ & $\frac12,\frac{ 2}{3} $ & $ 0$ & $\T_{ 7}^{\rm F},\cF(C_3)_2$   &  $  0 $  \\
		\hline
	
		\end{tabular}
			\caption{Fermionic Hecke images of $(SLY)_2$.}
			\label{tb:SLY2}
		\end{table}

\subsection{Type $(SLY)_1^2$}
Consider the double product of $SM_{\rm eff}(8,2)$ theory which we denote as $(SLY)_1^2$. Clearly, the central charge $c=\frac{3}{2}$ and the NS weights are $0,(\frac14)_2,\frac12$. The conductor becomes $N=16$. The full $S$-matrix can be easily deduced from the one of $(SLY)_1$ in \eqref{eq:SSLY1}. Consider the fermionic Hecke operation on $(SLY)_1^2$. We find there  exist  two classes $\rho(\sigma_p)^{\rm F}$ for $p\!\!\mod 8$, 
\be
\rho^{\rm F}(\sigma_{1,7})={\rm Id},\qquad \rho^{\rm F}(\sigma_{3,5})=\left(
\begin{array}{cccc}
 0 & 0 & 0 & 1 \\
 0 & 0 & -1 & 0 \\
 0 & -1 & 0 & 0 \\
 1 & 0 & 0 & 0 \\
\end{array}
\right) .
\ee
Obviously, they form a $\IZ_2$ group. 
We compute all fermionic Hecke images $\TF_p$ up to $p< 16$ and summarize those with $p<8$ in $c=12$ pairs in Table \ref{tb:SLY1square}. Interestingly, we find the $\TF_5$ image describes exactly the $SU(4)_4/\mathbb{Z}_2$ theory studied in \cite{Bae:2021lvk}. This orbifold theory has unbroken supersymmetry. See \cite[Equation (3.19)]{Bae:2021lvk} for the relation between fermionic characters and affine characters. Besides, the $\TF_3$ image describes $ \cF(A_1)_6^2$ as inherited from the fermionic Hecke operation of $(SLY)_1$ discussed in Section \ref{sec:SLY1}.
\begin{table}[ht]
\def\arraystretch{1.1}
	\centering
	\begin{tabular}{|c|c|c|c|c|c|c|c|c|c|c|c|c|c|}
		\hline
		 $c$ & $h_{\rm NS}$  & $m_{1/2}$ & remark & $\tc$ & $\tilh_{\rm NS}$ &  $\tm_{1/2}$ & remark & $K(\tau)+n$ \\
		\hline
$\frac32$ & $ (\frac14)_2,\frac12 $ & $2  $ & $\T_{1}^{\rm F}, (SLY)_1^2 $ & $ \frac{21}{2}$ & $ \frac12,(\frac34)_2 $ & $14  $ & $\T_{ 7}^{\rm F} $   &  $ 16  $  \\
$\frac92$ & $ (\frac14)_2,\frac12$ & $ 0$ & $\T_{3}^{\rm F}, \cF(A_1)_6^2 $ & $\frac{15}{2}$ & $  \frac12,(\frac34)_2 $ & $ 0$ & $\T_{ 5}^{\rm F}, \cF(A_3)_4$   &  $  0 $  \\
		\hline
	
		\end{tabular}
			\caption{Fermionic Hecke images of $(SLY)_1^2$.}
			\label{tb:SLY1square}
		\end{table}

\subsection{Type Ising$\,\otimes F$}
As we reviewed earlier, the critical Ising model can be fermionized to a free chiral fermion $F$. Let us consider the double product of the Ising model, but only fermionize one of them. Obviously, we obtain a $c=1$ fermionic RCFT with $h_{\rm NS}=0,\frac12,\frac{1}{16}$. This is a unitary theory appearing in the rank-6 SMC. 
The $S$-matrix of Ising model can be found in e.g. \cite[Equation (4.2)]{Duan:2022ltz}. Coupling with a free fermion does not change the $S$-matrix. 
The conductor also remains the same as $N=48$.

The bosonic Hecke images of Ising model were studied in \cite{Harvey:2018rdc}, see also \cite[Section 4.1]{Duan:2022ltz}, while the bosonic Hecke images of Ising$^2$ model were studied in  \cite[Section 7.7]{Duan:2022ltz}. The current study on the fermionic Hecke images of Ising$\,\otimes F$ is somewhat between them. We find there exist in total two classes of $\rh$ for $p\!\!\mod 24$: for $p\equiv 1, 7, 17, 23  \!\!\mod24$, i.e., $p^2\equiv1\!\!\mod24$, $\rh=\rm Id$, for $p\equiv5, 11, 13, 19\!\!\mod24$, i.e., $p^2\equiv25\!\!\mod24$, 
    \be
    \ba
\rho^{\rm F}(\sigma_{p})= \left(
\begin{array}{ccc}
 0 & 1 & 0 \\
 1 & 0 & 0 \\
 0 & 0 & -1 \\
\end{array}
\right) .
\ea
\ee
Not surprisingly, these are just the same with the two bosonic Hecke classes for Ising model given in \cite[Section 4.1]{Duan:2022ltz}. We find the fermionic Hecke images $\TF_p (\textrm{Ising}\,\otimes F)$ for $p<12$ is just the bosonic Hecke images $\T_p (\textrm{Ising})$ coupling with $p$ number of fermions. One can also introduce the generalized fermionic Hecke relations for $p=3k,k\in\mathbb{Z}$, where $\TF_3$ is defined by $(A_1)_2\otimes F^3$. We compute all fermionic Hecke images $\TF_p$ for $p<24$ and summarize them in $c=24$ pairs in Table \ref{tb:IsingF}. Note in this case, it is not possible to pair two $\TF_p$ images as a $c=12$ holomorphic theory. Nevertheless, for $p_1+p_2=8$, two images  $\TF_{p_1}$ and $\TF_{p_2}$ can form a holomorphic SCFT of 16 free chiral fermions. 
\begin{table}[ht]
\def\arraystretch{1.1}
	\centering
	\begin{tabular}{|c|c|c|c|c|c|c|c|c|c|c|c|c|c|}
		\hline
		 $c$ & $h_{\rm NS}$  & $m_{1/2}$ & remark & $\tc$ & $\tilh_{\rm NS}$ &  $\tm_{1/2}$ & remark   \\
		\hline
$1$ & $ \frac{1}{16},\frac12  $ & $ 1 $ & $\T_{1}^{\rm F}, \textrm{Ising}\,\otimes F  $ & $ 23$ & $1,\frac{23}{16}   $ & $ 0 $ & $\T_{ 23}^{\rm F},2\cdot 2^{1+22}Co_2 $      \\
$3$ & $ \frac{3}{16},\frac12 $ & $ 3 $ & $\T_{3}^{\rm F}, (A_1)_2\otimes F^3  $ &  $ 21$ & $ 1,\frac{21}{16}  $ & $ 0 $ & $\T_{21}^{\rm F}, $     \\
$5$ & $ \frac{5}{16},\frac12 $ & $5  $ & $\T_{5}^{\rm F}, (B_2)_1\otimes F^5  $ & $ 19$ & $ 1,\frac{19}{16}  $ & $ 0 $ & $\T_{ 19}^{\rm F}, $ \\
$7$ & $  \frac{7}{16},\frac12  $ & $ 7$ & $\T_{ 7}^{\rm F},(B_3)_1\otimes F^7 $ &  $ 17$ & $1,\frac{17}{16}   $ & $ 0 $ & $\T_{ 17}^{\rm F}, $ \\
$9$  &  $\frac12,\frac{9}{16}$ & $9$ & $\T_{ 9}^{\rm F},(B_4)_1\otimes F^9 $   & $15$ & $\frac{15}{16},1$ & $ 0$ & $\T_{ 15}^{\rm F} $   \\
$11$  &  $\frac12,\frac{11}{16}$ & $11$ & $\T_{ 11}^{\rm F} ,(B_5)_1\otimes F^{11}$   &  $13$  & $\frac{13}{16},1$ & $0$ & $\T_{ 13}^{\rm F} $  \\
		\hline
	
		\end{tabular}
			\caption{(Generalized) fermionic Hecke images of Ising$\,\otimes F$.}
			\label{tb:IsingF}
		\end{table}
		
In the end, we remark that the $\TF_{23}$ image describes a SCFT associated to the second largest Conway group $Co_2$, or more precisely a multi-covering $2\cdot 2^{1+22}Co_2$. 
We compute the NS characters of $\TF_{23}$ image as
\be
\ba
\chi_0^{\TF_{23}}&=q^{-\frac{23}{24}}(1+2300 q^{3/2}+46851 q^2+529828 q^{5/2}+4310154 q^3+\dots),\\
\chi_1^{\TF_{23}}&=q^{\frac{1}{24}}(23+2300 \sqrt{q}+46598 q+529828 q^{3/2}+4311948 q^2+\dots),\\
\chi_{\frac{23}{16}}^{\TF_{23}}&=q^{\frac{23}{48}}(2048+47104 \sqrt{q}+565248 q+4757504 q^{3/2}+31700992 q^2+\dots).
\ea
\ee
One can easily recognize some irreducible representations $\bf 23,2048,2300, 47104,$... of $2\cdot 2^{1+22}Co_2$ from the Fourier coefficients. See a collection of the dimensions of the irreducible representations of $2\cdot 2^{1+22}Co_2$ in e.g. \cite{Bae:2018qfh}. This is not surprising as the bosonic Hecke image $\T_{23}$ of Ising$^2$ has been associated to $2\cdot 2^{1+22}Co_2$ in  \cite[Section 3.2.8]{Bae:2020pvv}.

\subsection{Type $SM_{\rm eff}(7,3)$}
Supersymmetric minimal model $SM(7,3)$ has $c=-\frac{11}{14}$ and $c_{\rm eff}=\frac{13}{14}$. The NS conformal weights are $h^{\rm NS}=0,-\frac{1}{14},\frac27$, while $h^{\rm NS}_{\rm eff}=0,\frac{1}{14},\frac{5}{14}$. The conductor $N=336$. It was noticed in \cite{Melzer:1994qp} that $SM(7,3)$ can be realized as the fermionization of the $E_6$ invariant of bosonic minimal model $M(12,7)$. See \cite[Equation (4.5)]{Melzer:1994qp} for the character relations. 
We find the $S$-matrix of the NS characters of $SM_{\rm eff}(7,3)$ to be
\be
S=\frac{2}{\sqrt{7}}\left(
\begin{array}{ccc}
 \cos \left(\frac{\pi }{14}\right) & \sin \left(\frac{\pi }{7}\right) & \cos \left(\frac{3 \pi }{14}\right) \\
 \sin \left(\frac{\pi }{7}\right) & \cos \left(\frac{3 \pi }{14}\right) & -\cos \left(\frac{\pi }{14}\right) \\
 \cos \left(\frac{3 \pi }{14}\right) & -\cos \left(\frac{\pi }{14}\right) & -\sin \left(\frac{\pi }{7}\right) \\
\end{array}
\right).
\ee
It is easy to check $S^2=\rm Id$. We remark that this is the same $S$-matrix as the bosonic minimal model $M_{\rm eff}(7,2)$ in the order of weights $0,\frac37,\frac17$, see e.g. \cite[Equation (4.13)]{Duan:2022ltz}. We find there exist in total three classes for the fermionic Hecke operation of $SM_{\rm eff}(7,3)$: for $p\equiv1, 13, 29, 41, 43, 55, 71, 83, 85, 97, 113, 125, 127, 139, 155, 167 \!\!\mod168$, i.e., $p^2\equiv1\!\!\mod168$, $\rh=\rm Id$, for $p\equiv5, 19, 23, 37, 47, 61, 65, 79, 89, 103, 107, 121, 131, 145, 149, 163 \!\!\mod168$, i.e., $p^2\equiv25\!\!\mod168$, 
    \be
    \ba
\rho^{\rm F}(\sigma_{p})= \left(
\begin{array}{ccc}
 0 & 0 & 1 \\
 -1 & 0 & 0 \\
 0 & -1 & 0 \\
\end{array}
\right),
\ea
\ee
finally, for $p\equiv11, 17, 25, 31, 53, 59, 67, 73, 95, 101, 109, 115, 137, 143, 151, 157 \!\!\mod168$, i.e., $p^2\equiv121\!\!\mod168$, 
    \be
    \ba
\rho^{\rm F}(\sigma_{p})= \left(
\begin{array}{ccc}
 0 & -1 & 0 \\
 0 & 0 & -1 \\
 1 & 0 & 0 \\
\end{array}
\right).
\ea
\ee
Clearly, the three fermionic Hecke classes form a $\IZ_3$ group.  We compute all fermionic Hecke images $\TF_p$ up to $p< 14$ and summarize them in Table \ref{tb:SM73}. Although we do not recognize any interesting fermionic Hecke images, it still interesting to see that all six objects in rank-6 SMC related to divisor 7 (see e.g. \cite[Table 2]{Cho:2022kzf}) can be generated by $\TF_p$ on $SM_{\rm eff}(7,3)$.

\begin{table}[ht]
\def\arraystretch{1.1}
	\centering
	\begin{tabular}{|c|c|c|c|c|c|c|c|c|c|c|c|c|c|}
		\hline
		 $c$ & $h_{\rm NS}$  & $m_{1/2}$ & remark  & SMC \\
		\hline
$\frac{13}{14}$ & $\frac{1}{14},\frac{5}{14} $ & $ 1 $ & $\T_{1}^{\rm F},SM_{\rm eff}(7,3) $ & $6^S_{-1/14}$ \\
$\frac{65}{14}$ & $\frac{2}{7},\frac{5}{14} $ & $ 5 $ & $\T_{5}^{\rm F},\star $ & $6^S_{1/7}$ \\
$\frac{143}{14}$ & $\frac{3}{7},\frac{11}{14} $ & $11  $ & $\T_{11}^{\rm F}  $ & $6^S_{3/14}$ \\
$\frac{169}{14}$ & $\frac{9}{14},\frac{13}{14}  $ & $ 13 $ & $\T_{13}^{\rm F}$ & $6^S_{1/14}$ \\
$\frac{221}{14}$ & $ \frac{5}{7},\frac{15}{14} $ & $ 0 $ & $\T_{17}^{\rm F} $  & $6^S_{-3/14}$\\
$\frac{247}{14}$ & $ \frac{11}{14},\frac{6}{7} $ & $  0$ & $\T_{19}^{\rm F},\star $ & $6^S_{1/7}$ \\
$\frac{299}{14}$ & $ \frac{8}{7},\frac{17}{14} $ & $  0$ & $\T_{23}^{\rm F},\star $  & $6^S_{-1/7}$\\
		\hline
	
		\end{tabular}
			\caption{Fermionic Hecke images of $SM_{\rm eff}(7,3)$.}
			\label{tb:SM73}
		\end{table}

\subsection{Type $SM_{\rm sub}(20,2)$}\label{SMsub202}
Supersymmetric minimal model $SM(20,2)$ has $c=-\frac{114}{5}$ and $c_{\rm eff}=\frac{6}{5}$. 
The NS weights are $h^{\rm NS}=0,-\frac{2}{5},-\frac{7}{10},-\frac{9}{10},-1$, while $h^{\rm NS}_{\rm eff}=0,\frac{1}{10},\frac{3}{10},\frac{3}{5},1$. This suggests $SM(20,2)$ is a degenerate theory. Let us consider a sub-theory of $SM_{\rm eff}(20,2)$, i.e., a D-type non-diagonal modular invariant composed of
\be
\ba
\chi_0&=\chi_{0}^{SM_{\rm eff}(20,2)}+\chi_{1}^{SM_{\rm eff}(20,2)}=q^{-\frac{1}{20}}(1+\sqrt{q}+2 q+2 q^{3/2}+3 q^2+\dots),\\
\chi_{1/10}&=\chi_{1/10}^{SM_{\rm eff}(20,2)}-\chi_{3/5}^{SM_{\rm eff}(20,2)}=q^{\frac{1}{20}}(1+q^{3/2}+2 q^2+2 q^{5/2}+2 q^3+\dots),\\
\chi_{3/10}&=\chi_{3/10}^{SM_{\rm eff}(20,2)}=q^{\frac14}(1+\sqrt{q}+q+2 q^{3/2}+3 q^2+3 q^{5/2}+4 q^3+\dots).
\ea
\ee
It is easy to check 
\be
Z_{\rm NS}=|\chi_0|^2+| \chi_{1/10}|^2+2|\chi_{3/10} |^2
\ee
is $\Gamma_{\theta}$ modular invariant. Therefore the weigth-$\frac{3}{10}$ NS primary has degeneracy 2. The extended $S$-matrix can be deduced from the full $S$-matrix of $SM_{\rm eff}(20,2)$ as 
\be
S=\sqrt{\frac{2}{5}}\left(
\begin{array}{cccc}
 \sin \left(\frac{\pi }{20}\right)+\cos \left(\frac{\pi }{20}\right) & \cos \left(\frac{3 \pi }{20}\right)-\sin \left(\frac{3 \pi }{20}\right) & \frac{1}{\sqrt{2}} & \frac{1}{\sqrt{2}} \\
 \cos \left(\frac{3 \pi }{20}\right)-\sin \left(\frac{3 \pi }{20}\right) & \sin \left(\frac{\pi }{20}\right)+\cos \left(\frac{\pi }{20}\right) & -\frac{1}{\sqrt{2}} & -\frac{1}{\sqrt{2}} \\
 \frac{1}{\sqrt{2}} & -\frac{1}{\sqrt{2}} & \frac{\alpha_-}{2} & -\frac{\alpha_+}{2}  \\
 \frac{1}{\sqrt{2}} & -\frac{1}{\sqrt{2}} & -\frac{\alpha_+}{2}  & \frac{\alpha_-}{2} \\
\end{array}
\right) .
\ee
Here $\alpha_\pm=\sqrt{3\pm \sqrt{5}}$. We denote this sub-theory of $SM_{\rm eff}(20,2)$ as $SM_{\rm sub}(20,2)$. Clearly the conductor $N=20$. 

Consider the fermionic Hecke operation on $SM_{\rm sub}(20,2)$. From the above $S$-matrix, we find  there exist two classes of $\rh$ for $p\!\!\mod 10$:
\be
\rho^{\rm F}(\sigma_{1,9})={\rm Id},\qquad \rho^{\rm F}(\sigma_{3,7})= \left(
\begin{array}{cccc}
 0 & 1 & 0 & 0 \\
 1 & 0 & 0 & 0 \\
 0 & 0 & 0 & -1 \\
 0 & 0 & -1 & 0 \\
\end{array}
\right).
\ee
Obviously, they form a $\IZ_2$ group. We compute all fermionic Hecke images $\TF_p$ for $p< 20$ and summarize them as $c=12$ pairs in Table \ref{tb:SM202}. Notably, we find the $\TF_3$ image describes exactly the fermionization of $SO(4)_3$, i.e. $SU(2)_3^2$. The relation between the NS characters of and the affine characters can be found in for example \cite[Equation (3.67)]{Duan:2022ltz}. Besides, the $\TF_7$ image gives exactly the fermionization of $Sp(6)_1^2$. The relation between the NS characters of and the affine characters can be found in for example \cite[Equation (3.78)]{Duan:2022ltz}. 
Moreover, we find an undetermined theory with $c=\frac{66}{5},h_i=\frac45,\frac{11}{10}$ bootstraped in \cite{Duan:2022ltz} as a BPS solution of 3rd order $\Gamma_\theta$ MLDE is exactly our $\TF_{11}$ image. This implies that the weight-$\frac45$ NS primary has degeneracy 2, and the $S$-matrix and the fusion rules of the theory is the same with those of $SM_{\rm sub}(20,2)$ itself. Therefore the $c=\frac{66}{5}$ theory if exists is non-unitary and cannot be realized as the fermionization of a WZW model.

\begin{table}[ht]
\def\arraystretch{1.1}
	\centering
	\begin{tabular}{|c|c|c|c|c|c|c|c|c|c|c|c|c|c|}
		\hline
		 $c$ & $h_{\rm NS}$  & $m_{1/2}$ & remark  &$\tc$ & $\tilh_{\rm NS}$ &  $\tm_{1/2}$ & remark & $\!K(\tau)+n\!$ \\
		\hline
$\frac65$ & $\frac{1}{10},(\frac{3}{10})_2$ & $ 1$ & $\TF_1,SM_{\rm sub}(20,2)$ & $\frac{54}{5}$ & $ (\frac{7}{10})_2,\frac{9}{10} $ & $ 9 $ & $\T_{ 9}^{\rm F} $ & $10$  \\
$\frac{18}{5}$ & $ \frac{3}{10},(\frac{2}{5})_2 $ & $0  $ & $\T_{ 3}^{\rm F},  \cF(D_2)_3$ & $\frac{42}{5}$ & $ (\frac{3}{5})_2,\frac{7}{10} $ & $  0$ & $\T_{7 }^{\rm F},\cF (C_3)_1^2 $ & $0$ \\
		\hline
		\end{tabular}
			\caption{Fermionic Hecke images of $SM_{\rm sub}(20,2)$.}
			\label{tb:SM202}
		\end{table}

As a side remark, we also studied the fermionic Hecke operation on the full $SM_{\rm eff}(20,2)$, i.e, $(SLY)_4$, as a theory with five NS primaries. Unfortunately, we did not find any interesting fermionic Hecke images.

\subsection{Type $\cF({\textrm{Ising}}^3)$}\label{sec:som13}
A large class of supersymmetric RCFTs is known as the fermionization of WZW $SO(m)_1^3$ theories \cite{Johnson-Freyd:2019wgb}. These theories have central charge $c=\frac{3m}{2}$ and NS weights with degeneracy $h_{\rm NS}=0,(\frac12)_3,(\frac{m}{8})_3$. Recall WZW $SO(m)_1$ model has conformal weights $0,\frac{m}{16},\frac12$. The NS characters of $\cF (SO(m)_1^3)$  is defined by
\be\label{chisom13}
\ba
\chi_0^{\cF (SO(m)_1^3) }&=(\chi_0^{SO(m)_1 })^3+(\chi_{1/2}^{SO(m)_1 })^3,\\
\chi_{1/2}^{\cF (SO(m)_1^3) }&=\chi_0^{SO(m)_1 }\chi_{1/2}^{SO(m)_1 }\big(\chi_0^{SO(m)_1 }+\chi_{1/2}^{SO(m)_1 }\big),\\
\chi_{m/8}^{\cF (SO(m)_1^3) }&=\big(\chi_{1/2}^{SO(m)_1 }\big)^2\big(\chi_0^{SO(m)_1 }+\chi_{1/2}^{SO(m)_1 }\big).
\ea
\ee
When $m=1$, as $SO(1)_1$ is just the Ising model, we denote the supersymmetric theory as $\cF({\textrm{Ising}}^3)$. Note it is different from $(\cF{\textrm{Ising}}^3)=3F$, or $\textrm{Ising}\otimes 2F$, or $\textrm{Ising}^2\otimes F$. The $S$-matrix of the three NS characters in \eqref{chisom13} can be easily determined from the one of $SO(m)_1$ as
\be\label{SIsing3}
S=\frac{1}{4}\left(
\begin{array}{ccc}
 1 & 3 & 6 \\
 1 & 3 & -2 \\
 2 & -2 & 0 \\
\end{array}
\right) .
\ee
Note it is independent from $m$. 
Considering the degeneracy, the $7\times7$ full $S$-matrix can be found in e.g. \cite[Equation (3.47)]{Bae:2021mej}. Both $S$-matrices can be used to compute fermionic Hecke images, here for simplicity we use the reduced one \eqref{SIsing3}. 

Consider the fermionic Hecke operation on $\cF({\textrm{Ising}}^3)$. The conductor $N=16$. From \eqref{SIsing3}, we find for arbitrary \emph{odd} $p$, $\rh=\rm Id$, i.e., there is only one fermionic Hecke class. By computing $\TF_p$ for all $p<16$, we find that for $p<8$ the $\TF_p$ images exactly describes $\cF SO(p)_1^3$. We summarize those with $p<8$ in $c=12$ pairs in Table \ref{tb:HeckeIsing3}. The bilinear relation of the NS characters of each pair equals to $K(\tau)$. For $8<p<16$, $\TF_p$ is different from $\cF SO(p)_1^3$. Nevertheless, similar with the one free fermion case, we find the NS characters of $\cF SO(p)_1^3$ can be written as the linear combinations of $\TF_{p}$ and $\TF_{p-8}$ images. For example, for $p=11$, we find the three NS characters of weights $0,\frac12,\frac{11}{8}$ can be written as
\be
\ba
\chi^{\cF SO(11)_1^3}&=\TF_{11}+11M\cdot \TF_3,\qquad M=\left(
\begin{array}{ccc}
 0 & 3 & 0 \\
 1 & 2 & 0 \\
 0 & 0 & -1 \\
\end{array}
\right).
\ea
\ee

\begin{table}[ht]
\def\arraystretch{1.1}
	\centering
	\begin{tabular}{|c|c|c|c|c|c|c|c|c|c|c|c|c|c|}
		\hline
		 $c$ & $h_{\rm NS}$  & $m_{1/2}$ & remark & $\tc$ & $\tilh_{\rm NS}$ &  $\tm_{1/2}$ & remark & $K(\tau)+n$ \\
		\hline
$\frac{3p}{2}$ & $(\frac12)_3,(\frac{p}{8})_3 $ & $0 $ & $\T_{p}^{\rm F} $ & $\frac{3(8-p)}{2}$ & $(\frac12)_3,(\frac{8-p}{8})_3 $ & $ 0$ & $\T_{16-p}^{\rm F}$   &  $  0 $  \\
		\hline
	
		\end{tabular}
			\caption{(Generalized) fermionic Hecke images of $\cF({\textrm{Ising}}^3)$.}
			\label{tb:HeckeIsing3}
		\end{table}
		
Now we would like to show that the above results can be safely extended to even $p$ by including generalized fermionic Hecke images $\TF_2$, $\TF_4$ and $\TF_8$ which describe $\cF SO(2)_1^3$, $\cF SO(4)_1^3$ and $\cF SO(8)_1^3$ respectively. This is very similar with the situation of one free chiral fermion in Section \ref{sec:1F}. Here we just show for generalized $\TF_2$. Denote the three NS characters of $\cF({\textrm{Ising}}^3)$ as $\chi_0,\chi_{1/2},\chi_{1/8}$. We find the three NS characters of $\cF SO(2)_1^3$ can be written as degree 2 polynomials of $\chi_0,\chi_{1/2},\chi_{1/8}$ as
\be
\ba
\chi_0^{\cF SO(2)_1^3}&=\chi_0^2+3\chi_{1/2}^2,\\
\chi_{1/2}^{\cF SO(2)_1^3}&=2\chi_0\chi_{1/2}+2\chi_{1/2}^2,\\
\chi_{1/4}^{\cF SO(2)_1^3}&=\chi_{1/8}^2.
\ea
\ee
This shows that ${\cF SO(2)_1^3}$ can be regarded as a generalized $\TF_2$ image of $\cF({\textrm{Ising}}^3)$. Then we can use the $\TF_k$ images of ${\cF SO(2)_1^3}$ to defined the generalized $\TF_{2k}$ images of $\cF({\textrm{Ising}}^3)$, with $k$ coprime to the new conductor $8$. There is still only one fermionic Hecke class $\rh=\rm Id$. This procedure can be further extended to $\cF SO(4)_1^3$ and $\cF SO(8)_1^3$. In summary, we find all information in Table \ref{tb:HeckeIsing3} still holds for all generalized $\TF_p$ images.

\section{Four NS characters}\label{sec:4chi}
\subsection{Type $(SLY)_3$}
The non-unitary supersymmetric minimal model $SM(16,2)$ has 
$c=-\frac{135}{8}$ and $h=0,-\frac{3}{8}$, $-\frac58,-\frac34$, while $SM_{\rm eff}(16,2)$, i.e., $(SLY)_3$ has $c_{\rm eff}=\frac{9}{8}$ and $h_{\rm eff}=0,\frac{1}{8},\frac38,\frac{3}{4}$. The conductor $N=64$. We find the $S$-matrix for $(SLY)_3$ is
\be
S=\frac{1}{\sqrt{2}}\left(
\begin{array}{cccc}
 \cos \left(\frac{\pi }{16}\right) & \cos \left(\frac{3 \pi }{16}\right) & \sin \left(\frac{3 \pi }{16}\right) & \sin \left(\frac{\pi }{16}\right) \\
 \cos \left(\frac{3 \pi }{16}\right) & -\sin \left(\frac{\pi }{16}\right) & -\cos \left(\frac{\pi }{16}\right) & -\sin \left(\frac{3 \pi }{16}\right) \\
 \sin \left(\frac{3 \pi }{16}\right) & -\cos \left(\frac{\pi }{16}\right) & \sin \left(\frac{\pi }{16}\right) & \cos \left(\frac{3 \pi }{16}\right) \\
 \sin \left(\frac{\pi }{16}\right) & -\sin \left(\frac{3 \pi }{16}\right) & \cos \left(\frac{3 \pi }{16}\right) & -\cos \left(\frac{\pi }{16}\right) \\
\end{array}
\right).
\ee
It is easy to check $S^2=\rm Id$.

Consider the fermionic Hecke operation of $(SLY)_3$. We find there exist in total 8 classes of $\rh$ for $p\!\!\mod 32$:  
\be\nonumber
\ba
\rho^{\rm F}(\sigma_{1,31})=-\rho^{\rm F}(\sigma_{15,17})={\rm Id},\qquad &\rho^{\rm F}(\sigma_{3,29})=-\rho^{\rm F}(\sigma_{13, 19})=\left(
\begin{array}{cccc}
 0 & 0 & 1 & 0 \\
 -1 & 0 & 0 & 0 \\
 0 & 0 & 0 & 1 \\
 0 & 1 & 0 & 0 \\
\end{array}
\right),\\
\rho^{\rm F}(\sigma_{5,27})=-\rho^{\rm F}(\sigma_{11,21})= \left(
\begin{array}{cccc}
 0 & 1 & 0 & 0 \\
 0 & 0 & 0 & -1 \\
 -1 & 0 & 0 & 0 \\
 0 & 0 & -1 & 0 \\
\end{array}
\right),\quad &
\rho^{\rm F}(\sigma_{7,25})=-\rho^{\rm F}(\sigma_{9,23})=\left(
\begin{array}{cccc}
 0 & 0 & 0 & -1 \\
 0 & 0 & 1 & 0 \\
 0 & -1 & 0 & 0 \\
 1 & 0 & 0 & 0 \\
\end{array}
\right)  .
\ea
\ee
We find the 8 fermionic Hecke classes form a $\IZ_8$ group.  We compute all fermionic Hecke images $\TF_p$ up to $p< 14$ and summarize the relevant information in Table \ref{tb:SLY3}. Notably, we find the $\TF_7$ image describes exactly the fermionization of WZW $SO(7)_3$ model. The relation between the NS characters of and the affine characters can be found in e.g. \cite[Equation (4.45)]{Duan:2022ltz}.

\begin{table}[ht]
\def\arraystretch{1.1}
	\centering
	\begin{tabular}{|c|c|c|c|c|c|c|c|c|c|c|c|c|c|}
		\hline
		 $c$ & $h_{\rm NS}$  & $m_{1/2}$ & remark   \\
		\hline
$\frac98$ & $ \frac{1}{8},\frac38,\frac{3}{4} $ & $  1$  &  $\TF_1,(SLY)_3$  \\
$\frac{27}{8}$ & $ \frac{1}{8},\frac{1}{4},\frac{3}{8} $ & $ 3 $  &  $\TF_3,\star$  \\
$\frac{45}{8}$ & $ \frac{1}{4},\frac{3}{8},\frac{5}{8} $ & $ 5 $  &  $\TF_5,\star$  \\
$\frac{63}{8}$ & $ \frac{3}{8},\frac{5}{8},\frac{3}{4} $ & $ 0 $  &  $\TF_7,\cF(B_3)_3$  \\
$\frac{81}{8}$ & $ \frac{3}{8},\frac{5}{8},\frac{3}{4} $ & $ 9 $  &  $\TF_9,\star$  \\
$\frac{99}{8}$ & $ \frac{3}{4},\frac{7}{8},\frac{9}{8} $ & $ 0 $  &  $\TF_{11}$  \\
$\frac{117}{8}$ & $  \frac{3}{4},\frac{7}{8},\frac{9}{8}$ & $0  $  &  $\TF_{13}$  \\
		\hline
	
		\end{tabular}
			\caption{Fermionic Hecke images of $(SLY)_3$.}
			\label{tb:SLY3}
		\end{table}

\subsection{Type $SM(6,4)$}
The supersymmetric minimal model $SM(6,4)$ is a degenerate unitary  theory with $c=1$. 
There are four NS primaries with weights $0,\frac{1}{16},\frac16,1$. It is easy to find the $S$-matrix is
\be
S=\frac{1}{2\sqrt{3}}\left(
\begin{array}{cccc}
 1 & \sqrt{6} & 2 & 1 \\
 \sqrt{6} & 0 & 0 & -\sqrt{6} \\
 2 & 0 & -2 & 2 \\
 1 & -\sqrt{6} & 2 & 1 \\
\end{array}
\right) .
\ee
In Section \ref{sec:S2A1}, we have discussed a sub-theory of $SM(6,4)$ that is $\cS^2A_1$. Note here the conductor $N=48$ is bigger than its sub-theory.

Consider the fermionic Hecke operation of $SM(6,4)$. We find there exist four classes of $\rh$ for $p\!\!\mod 24$: 
\be\nonumber
\ba
\rho^{\rm F}(\sigma_{1,23})=-\rho^{\rm F}(\sigma_{5,19})={\rm Id},\qquad &\rho^{\rm F}(\sigma_{7,17})=-\rho^{\rm F}(\sigma_{11, 13})=\left(
\begin{array}{cccc}
 0 & 0 & 0 & -1 \\
 0 & 1 & 0 & 0 \\
 0 & 0 & -1 & 0 \\
 -1 & 0 & 0 & 0 \\
\end{array}
\right).
\ea
\ee
We compute all fermionic Hecke images $\TF_p$ up to $p<24$ and summarize them in $c=24$ pairs in Table \ref{tb:SM64}. 
It is worthwhile to point out that the $\TF_{11}$ image of $SM(6,4)$ describes the fermionization of $(D_6)_2$, while  the $\TF_{11}$ image of the sub-theory of $SM(6,4)$ describes the fermionization of $SU(12)_1$. For the precise relation between the fermionic characters of $(D_6)_2$ and the affine characters, we refer to \cite[Equation (3.27)]{Bae:2021lvk}. It is easy to check the characters satisfy the bilinear relations
$
\TF_{11}\cdot \TF_{13}=K(\tau)^2-408={q}^{-1}+144+\dots.
$ 
Similarly, we find $\TF_{1}\cdot \TF_{23}=K(\tau)^2-552$.

\begin{table}[ht]
\def\arraystretch{1.1}
	\centering
	\begin{tabular}{|c|c|c|c|c|c|c|c|c|c|c|c|c|c|}
		\hline
		 $c$ & $h_{\rm NS}$  & $m_{1/2}$ & remark & $\tc$ & $\tilh_{\rm NS}$ &  $\tm_{1/2}$ & remark   \\
		\hline
$1$ &  $ \frac{1}{16},\frac16,1$ & $ 0 $ & $\T_{1}^{\rm F},  SM(6,4)  $ & $ 23$ & $ 1,\frac{4}{3},\frac{23}{16}   $ & $ 0 $ & $\T_{23}^{\rm F}  $      \\
$ 5$ & $\frac{5}{16},\frac{1}{3},\frac{1}{2} $ & $ 5 $ & $\T_{5 }^{\rm F}, \star   $ & $ 19$ & $ 1,\frac{7}{6},\frac{19}{16}   $ & $ 0 $ & $\T_{19 }^{\rm F},\star  $      \\
$7 $ &  $\frac{7}{16},\frac{1}{2},\frac{2}{3} $ & $ 7 $ & $\T_{7 }^{\rm F}, \star   $ & $17 $ & $   \frac{5}{6},1,\frac{17}{16} $ & $ 0 $ & $\T_{17 }^{\rm F},\star  $      \\
$ 11$ &  $ \frac{11}{16},\frac{5}{6},1$ & $ 0 $ & $\T_{11 }^{\rm F}, \cF(D_6)_2   $ & $ 13$ & $ \frac{2}{3},\frac{13}{16},1   $ & $ 0 $ & $\T_{ 13}^{\rm F}  $      \\
		\hline
	
		\end{tabular}
			\caption{Fermionic Hecke images of $SM(6,4)$.}
			\label{tb:SM64}
		\end{table}

\subsection{Type $SM_{\rm sub}(28,2)$}
Supersymmetric minimal model $SM(28,2)$ has central charge 
$c=-\frac{243}{7}$ and NS weights 
$h=0,-\frac{3}{7},-\frac{11}{14},-\frac{15}{14},-\frac{9}{7},-\frac{10}{7},-\frac{3}{2} $, while $SM_{\rm eff}(28,2)$ has $c_{\rm eff}=\frac{9}{7}$ and $h_{\rm eff}=0,\frac{1}{14},\frac{3}{14},\frac{3}{7},\frac{5}{7},\frac{15}{14},\frac{3}{2} $. The conductor $N=56$. Let us consider a sub-theory of $SM_{\rm eff}(28,2)$, i.e. a D-type non-diagonal modular invariant by
\be
\ba
\chi_0&=\chi_{0}^{SM_{\rm eff}(28,2)}-\chi_{3/2}^{SM_{\rm eff}(28,2)}=q^{-\frac{3}{56}}(1+\sqrt{q}+q+q^{3/2}+3 q^2+4 q^{5/2}+\dots),\\
\chi_{1/14}&=\chi_{1/14}^{SM_{\rm eff}(28,2)}+\chi_{15/14}^{SM_{\rm eff}(28,2)}=q^{\frac{1}{56}}(1+\sqrt{q}+2 q+3 q^{3/2}+4 q^2+5 q^{5/2}+\dots),\\
\chi_{3/14}&=\chi_{3/14}^{SM_{\rm eff}(28,2)}-\chi_{5/7}^{SM_{\rm eff}(28,2)}=q^{\frac{9}{56}}(1+q^{3/2}+q^2+q^{5/2}+2 q^3+3 q^{7/2}+\dots),\\
\chi_{3/7}&=\chi_{3/7}^{SM_{\rm eff}(28,2)}=q^{\frac{3}{8}}(1+\sqrt{q}+q+2 q^{3/2}+3 q^2+4 q^{5/2}+5 q^3+6 q^{7/2}+\dots).
\ea
\ee
It is easy to check 
\be
Z_{\rm NS}=|\chi_0|^2+| \chi_{1/14}|^2+|\chi_{3/14} |^2+2|\chi_{3/7} |^2
\ee
is $\Gamma_{\theta}$ modular invariant. Clearly the weight-$\frac{3}{7}$ primary has degeneracy 2. The extended $S$-matrix can be deduced from the full $S$-matrix of $SM_{\rm eff}(28,2)$ as 
\be\nonumber
S=\sqrt{\frac{2}{7}}\left(
\begin{array}{ccccc}
 \cos \left(\frac{\pi }{28}\right)-\sin \left(\frac{\pi }{28}\right) & \sin \left(\frac{3 \pi }{28}\right)+\cos \left(\frac{3 \pi }{28}\right) & \cos \left(\frac{5 \pi }{28}\right)-\sin \left(\frac{5 \pi }{28}\right) & \frac{1}{\sqrt{2}} & \frac{1}{\sqrt{2}} \\
 \sin \left(\frac{3 \pi }{28}\right)+\cos \left(\frac{3 \pi }{28}\right) & \sin \left(\frac{5 \pi }{28}\right)-\cos \left(\frac{5 \pi }{28}\right) & \cos \left(\frac{\pi }{28}\right)-\sin \left(\frac{\pi }{28}\right) & -\frac{1}{\sqrt{2}} & -\frac{1}{\sqrt{2}} \\
 \cos \left(\frac{5 \pi }{28}\right)-\sin \left(\frac{5 \pi }{28}\right) & \cos \left(\frac{\pi }{28}\right)-\sin \left(\frac{\pi }{28}\right) & -\sin \left(\frac{3 \pi }{28}\right)-\cos \left(\frac{3 \pi }{28}\right) & -\frac{1}{\sqrt{2}} & -\frac{1}{\sqrt{2}} \\
 \frac{1}{\sqrt{2}} & -\frac{1}{\sqrt{2}} & -\frac{1}{\sqrt{2}} & \frac{{\alpha_+}}{2} & -\frac{\alpha_-}{2} \\
 \frac{1}{\sqrt{2}} & -\frac{1}{\sqrt{2}} & -\frac{1}{\sqrt{2}} & -\frac{\alpha_-}{2}  & \frac{\alpha_+}{2} \\
\end{array}
\right) .
\ee
Here $\alpha_\pm=\sqrt{4\pm\sqrt{7}}$. One can check $S^2=\rm Id$. We denote this sub-theory as $SM_{\rm sub}(28,2)$. 

Consider the fermionic Hecke operation of $SM_{\rm sub}(28,2)$. We find there exist in total 6 classes of $\rh$ for $p\!\!\mod 28$: $\rho(\sigma_{1,27})=-\rho(\sigma_{13,15})=\rm Id$,
\be\nonumber
\ba
\rho^{\rm F}(\sigma_{3,25})=-\rho^{\rm F}(\sigma_{11,17})=\!
\left(
\begin{array}{ccccc}
 0 & 0 & -1 & 0 & 0 \\
 -1 & 0 & 0 & 0 & 0 \\
 0 & 1 & 0 & 0 & 0 \\
 0 & 0 & 0 & 1 & 0 \\
 0 & 0 & 0 & 0 & 1 \\
\end{array}
\right)
\!, \rho^{\rm F}(\sigma_{5,23})=-\rho^{\rm F}(\sigma_{9, 19})=\!\left(
\begin{array}{ccccc}
 0 & 1 & 0 & 0 & 0 \\
 0 & 0 & -1 & 0 & 0 \\
 1 & 0 & 0 & 0 & 0 \\
 0 & 0 & 0 & 0 & -1 \\
 0 & 0 & 0 & -1 & 0 \\
\end{array}
\right) \!.
\ea
\ee
They form a $\IZ_6$ abelian group, where the group elements $0,1,2,3,4,5$ are represented by $p=1,5,3,13,9,11$ respectively. We compute all $\TF_p$ for $p< 18$ and summarize the relevant data in Table \ref{tb:SM282sub}. 
Notably, we find the $\TF_5$ image describes exactly the fermionization of WZW $SO(6)_3$ model. The relation between the NS characters of and the affine characters can be found in for example \cite[Equation (4.33)]{Duan:2022ltz}.
\begin{table}[ht]
\def\arraystretch{1.1}
	\centering
	\begin{tabular}{|c|c|c|c|c|c|c|c|c|c|c|c|c|c|}
		\hline
		 $c$ & $h_{\rm NS}$  & $m_{1/2}$ & remark   \\
		\hline
$\frac97$ & $ \frac{1}{14},\frac{3}{14},(\frac{3}{7})_2 $ & $  1$  &  $\TF_1,SM_{\rm sub}(28,2)$  \\
$\frac{27}{7}$ & $ \frac{1}{7},\frac{3}{14},(\frac{2}{7})_2 $ & $ 6 $  &  $\TF_3,\star $  \\
$\frac{45}{7}$ & $ \frac{5}{14},\frac{4}{7},\frac{9}{14} $ & $  0$  &  $\TF_5,\cF(D_3)_3 $  \\
$\frac{81}{7}$ & $ \frac{3}{7},\frac{9}{14},\frac{6}{7},\frac{6}{7} $ & $ 9 $  &  $\TF_{9},\star $  \\
$\frac{99}{7}$ & $ (\frac{5}{7})_2,\frac{11}{14},\frac{6}{7} $ & $ 0 $  &  $\TF_{11} $  \\
$\frac{117}{7}$ & $ \frac{13}{14},(\frac{15}{14})_2,\frac{9}{7} $ & $ 0 $  &  $\TF_{13},\star $  \\
$\frac{135}{7}$ & $ \frac{5}{7},(\frac{13}{14})_2,\frac{15}{14} $ & $0  $  &  $\TF_{15},\star $  \\
$\frac{153}{7}$ & $ \frac{8}{7},\frac{17}{14},(\frac{9}{7})_2 $ & $ 0 $  &  $\TF_{17} $  \\
		\hline
	
		\end{tabular}
			\caption{Fermionic Hecke images of $SM_{\rm sub}(28,2)$.}
			\label{tb:SM282sub}
		\end{table}

\subsection{Type $SM_{\rm sub}(8,6)$} 
Unitary superysmmetric minimal model  $SM(8,6)$ has $c=\frac54$ and nine NS primaries with weights $h_{\rm NS}=0,\frac{1}{32},\frac{1}{12},\frac{5}{32},\frac{1}{4},\frac{5}{6},\frac{33}{32},\frac{5}{4},3$. To produce interesting fermionic Hecke images, let us consider the following non-diagonal modular invariant of $SM(8,6)$ composed of
\be\label{chiSM86}
\ba
\chi_0&=\chi_{0}^{SM(8,6)}+\chi_{3}^{SM(8,6)}=q^{-\frac{5}{96}}(1+q^{3/2}+q^2+q^{5/2}+2 q^3+3 q^{7/2}+\dots),\\
\chi_{1/12}&=\chi_{1/12}^{SM(8,6)}=q^{\frac{1}{32}}(1+\sqrt{q}+q+2 q^{3/2}+3 q^2+4 q^{5/2}+5 q^3+7 q^{7/2} +\dots),\\
\chi_{1/4}&=\chi_{1/4}^{SM(8,6)}+\chi_{5/4}^{SM(8,6)}=q^{\frac{19}{96}}(1+\sqrt{q}+2 q+2 q^{3/2}+3 q^2+5 q^{5/2}+\dots),\\
\chi_{5/6}&=\chi_{5/6}^{SM(8,6)}=q^{\frac{25}{32}}(1+\sqrt{q}+q+q^{3/2}+2 q^2+3 q^{5/2}+3 q^3+4 q^{7/2}+\dots).
\ea
\ee
It is easy to check 
\be
Z_{\rm NS}=|\chi_0|^2+2| \chi_{1/12}|^2+|\chi_{1/4} |^2+2|\chi_{5/6} |^2
\ee
is $\Gamma_{\theta}$ modular invariant. The weight-$\frac{1}{2}$ and $\frac56$ NS primaries both have degeneracy 2. This $\Gamma_{\theta}$ modular invariant has been studied in \cite{Kastor:1986ig,Matsuo:1986vc,Cappelli:1986ed}. Let us denote this sub-theory as $ SM_{\rm sub}(8,6)$. Obviously the conductor $N=96$. In fact, it can be also realized as a supersymmetric coset
\be
SM_{\rm sub}(8,6)=\frac{\cF(A_1)_6}{SM_{\rm sub}(6,4)}.
\ee
The character relations of this coset can be found in e.g. \cite[Equation (5.51)]{Bae:2020xzl}. The $S$-matrix of the four characters in \eqref{chiSM86} can be determined as
\be
S=\frac{1}{\sqrt{3}} \left(
\begin{array}{cccc}
 \sin \left(\frac{\pi }{8}\right) & 2 \cos \left(\frac{\pi }{8}\right) & \cos \left(\frac{\pi }{8}\right) & 2 \sin \left(\frac{\pi }{8}\right) \\
 \cos \left(\frac{\pi }{8}\right) & \sin \left(\frac{\pi }{8}\right) & -\sin \left(\frac{\pi }{8}\right) & -\cos \left(\frac{\pi }{8}\right) \\
 \cos \left(\frac{\pi }{8}\right) & -2 \sin \left(\frac{\pi }{8}\right) & -\sin \left(\frac{\pi }{8}\right) & 2 \cos \left(\frac{\pi }{8}\right) \\
 \sin \left(\frac{\pi }{8}\right) & -\cos \left(\frac{\pi }{8}\right) & \cos \left(\frac{\pi }{8}\right) & -\sin \left(\frac{\pi }{8}\right) \\
\end{array}
\right).
\ee
To lighten the computation, we use this non-symmetric reduced $S$-matrix instead of the  $6\times6$ full $S$-matrix.

Let us consider the fermionic Hecke operation on $SM_{\rm sub}(8,6)$. We find there exist in total 8 classes of $\rh$ for $p\!\!\mod 48$: $\rho^{\rm F}(\sigma_{1,7,41,47})=-\rho^{\rm F}(\sigma_{17, 23, 25, 31})=\rm Id$, while
\be
\rho^{\rm F}(\sigma_{5, 13, 35,43 })=-\rho^{\rm F}(\sigma_{11, 19, 29,37})=\left(
\begin{array}{cccc}
 0 & 0 & 1 & 0 \\
 0 & 0 & 0 & -1 \\
 -1 & 0 & 0 & 0 \\
 0 & 1 & 0 & 0 \\
\end{array}
\right).
\ee
The four Hecke classes form a $\IZ_4$ abelian group. We compute all fermionic Hecke images $\TF_p$ for $p<20$ and list the results in Table \ref{tb:SM86sub}. Notably, we find the $\TF_{7}$ image describes the renowned fermionization  of WZW $SU(6)_2$ theory. This theory has emergent supersymmetry and true supersymmetric vacua. The relation between the NS characters of $\cF(A_5)_2$ and the affine characters can be found in e.g. \cite[Table 9]{Bae:2021lvk}. 
\begin{table}[ht]
\def\arraystretch{1.1}
	\centering
	\begin{tabular}{|c|c|c|c|c|c|c|c|c|c|c|c|c|c|}
		\hline
		 $c$ & $h_{\rm NS}$  & $m_{1/2}$ & remark   \\
		\hline
$\frac{5}{4}$ & $(\frac{1}{12})_2,\frac14,(\frac56)_2  $ & $ 0 $  &  $\TF_{1},SM_{\rm sub}(8,6) $  \\
$\frac{25}{4}$ & $ \frac{1}{4},(\frac{5}{12})_2,(\frac{2}{3})_2 $ & $ 10 $  &  $\TF_{5},\star $  \\
$\frac{35}{4}$ & $ (\frac{7}{12})_2,\frac{3}{4},(\frac{5}{6})_2 $ & $ 0 $  &  $\TF_{7},\cF(A_5)_2 $  \\
$\frac{55}{4}$ & $ (\frac{2}{3})_2,\frac{3}{4},(\frac{11}{12})_2 $ & $ 0 $  &  $\TF_{11} $  \\
$\frac{65}{4}$ & $ (\frac{5}{6})_2,(\frac{13}{12})_2,\frac{5}{4} $ & $0  $  &  $\TF_{13},\star $  \\
$\frac{85}{4}$ & $(\frac{11}{12})_2,(\frac{7}{6})_2,\frac{5}{4}  $ & $ 0 $  &  $\TF_{17},\star $  \\
$\frac{95}{4}$ & $ (\frac{13}{12})_2,\frac{5}{4},(\frac{4}{3})_2 $ & $ 0 $  &  $\TF_{19} $  \\
		\hline
	
		\end{tabular}
			\caption{Fermionic Hecke images of $SM_{\rm sub}(8,6)$.}
			\label{tb:SM86sub}
		\end{table}

\section{Five NS characters}\label{sec:5chi}
\subsection{Type $SM_{\rm sub}(36,2)$}
Supersymmetric minimal model $SM(36,2)$ has central charge 
$c=-\frac{140}{3}$ and NS weights 
$h= 0,-\frac{4}{9},-\frac{5}{6},-\frac{7}{6},-\frac{13}{9},-\frac{5}{3},-\frac{11}{6},-\frac{35}{18},-2$, while the effective theory $SM_{\rm eff}(36,2)$ has $c_{\rm eff}=\frac{4}{3}$ and $h_{\rm eff}= 0,\frac{1}{18}$, $\frac{1}{6},\frac{1}{3},\frac{5}{9},\frac{5}{6},\frac{7}{6},\frac{14}{9},2 $.  Let us consider a sub-theory of $SM_{\rm eff}(36,2)$, that is a D-type non-diagonal modular invariant composed by
\be
\ba
\chi_0&=\chi_{0}^{SM_{\rm eff}(36,2)}+\chi_{2}^{SM_{\rm eff}(36,2)}=q^{-\frac{1}{18}}(1+\sqrt{q}+q+2 q^{3/2}+4 q^2+4 q^{5/2}+\dots),\\
\chi_{1/18}&=\chi_{1/18}^{SM_{\rm eff}(36,2)}-\chi_{14/9}^{SM_{\rm eff}(36,2)}=1+\sqrt{q}+q+q^{3/2}+2 q^2+3 q^{5/2}+4 q^3+\dots,\\
\chi_{1/6}&=\chi_{3/14}^{SM_{\rm eff}(36,2)}+\chi_{7/6}^{SM_{\rm eff}(36,2)}=q^{\frac{1}{9}}(1+\sqrt{q}+2 q+3 q^{3/2}+4 q^2+6 q^{5/2}+\dots),\\
\chi_{1/3}&=\chi_{1/3}^{SM_{\rm eff}(36,2)}-\chi_{5/6}^{SM_{\rm eff}(36,2)}=q^{\frac{5}{18}}(1+q^{3/2}+q^2+q^{5/2}+q^3+2 q^{7/2}+\dots),\\
\chi_{5/9}&=\chi_{5/9}^{SM_{\rm eff}(36,2)}=q^{\frac{1}{2}}(1+\sqrt{q}+q+2 q^{3/2}+3 q^2+4 q^{5/2}+5 q^3+7 q^{7/2}+\dots).
\ea
\ee
It is easy to check 
\be
Z_{\rm NS}=|\chi_0|^2+| \chi_{1/18}|^2+|\chi_{1/6} |^2+|\chi_{1/3} |^2+2|\chi_{5/9} |^2
\ee
is $\Gamma_{\theta}$ modular invariant. Clearly, the weight-$\frac59$ NS character has degeneracy 2. The extended $S$-matrix can be deduced from the full $S$-matrix of $SM_{\rm eff}(36,2)$ as 
\be\nonumber
\frac13\left(
\begin{array}{cccccc}
 \sqrt{2} \left(\sin (\frac{\pi }{36})+\cos (\frac{\pi }{36})\right) & 1 & \sqrt{2} \left(\sin (\frac{5 \pi }{36})+\cos (\frac{5 \pi }{36})\right) & \sqrt{2} \left(\cos (\frac{7 \pi }{36})-\sin (\frac{7 \pi }{36})\right) & 1 & 1 \\
 1 & 2 & -1 & 1 & -1 & -1 \\
 \sqrt{2} \left(\sin (\frac{5 \pi }{36})+\cos (\frac{5 \pi }{36})\right) & -1 & \sqrt{2} \left(\cos (\frac{7 \pi }{36})-\sin (\frac{7 \pi }{36})\right) & -\sqrt{2} \left(\sin (\frac{\pi }{36})+\cos (\frac{\pi }{36})\right) & -1 & -1 \\
 \sqrt{2} \left(\cos (\frac{7 \pi }{36})-\sin (\frac{7 \pi }{36})\right) & 1 & -\sqrt{2} \left(\sin (\frac{\pi }{36})+\cos (\frac{\pi }{36})\right) & -\sqrt{2} \left(\sin (\frac{5 \pi }{36})+\cos (\frac{5 \pi }{36})\right) & 1 & 1 \\
 1 & -1 & -1 & 1 & 2 & -1 \\
 1 & -1 & -1 & 1 & -1 & 2 \\
\end{array}
\right) .
\ee
We denote this sub-theory as $SM_{\rm sub}(36,2)$. Clearly the conductor $N=18$.
Note this is a degenerate theory.

Consider the fermionic Hecke operation on $SM_{\rm sub}(36,2)$. We find there exist three classes of $\rh$ for $p\!\!\mod 18$: $\rho^{\rm F}(\sigma_{1,17})=\rm Id$,
$$
\rho^{\rm F}(\sigma_{5,13})=\left(
\begin{array}{cccccc}
 0 & 0 & 0 & 1 & 0 & 0 \\
 0 & 1 & 0 & 0 & 0 & 0 \\
 -1 & 0 & 0 & 0 & 0 & 0 \\
 0 & 0 & -1 & 0 & 0 & 0 \\
 0 & 0 & 0 & 0 & 1 & 0 \\
 0 & 0 & 0 & 0 & 0 & 1 \\
\end{array}
\right),\quad \text{and}\quad
\rho^{\rm F}(\sigma_{7,11})=\left(
\begin{array}{cccccc}
 0 & 0 & -1 & 0 & 0 & 0 \\
 0 & 1 & 0 & 0 & 0 & 0 \\
 0 & 0 & 0 & -1 & 0 & 0 \\
 1 & 0 & 0 & 0 & 0 & 0 \\
 0 & 0 & 0 & 0 & 1 & 0 \\
 0 & 0 & 0 & 0 & 0 & 1 \\
\end{array}
\right) .
$$
They form a $\IZ_3$ abelian group. We compute all $\TF_p$ images for $p<18$ and summarize the results in $c=24$ pairs in Table \ref{tb:SM362sub}. 
Notably, we find the $\TF_7$ image describes exactly the fermionization of WZW $SO(8)_3$ model:
\be
\ba
\chi_0^{\TF_7}&= \chi_{0,0}^{SO(8)_3} +\chi_{\frac32,3w_1}^{SO(8)_3}=q^{-\frac{7}{18}}(1+28 q+112 q^{3/2}+434 q^2+1568 q^{5/2}+\dots)   ,\\
\chi_{\frac{7}{18}}^{\TF_7}&= \chi_{\frac{7}{18},w_1}^{SO(8)_3} +\chi_{\frac89,2w_1}^{SO(8)_3}=  8+35 \sqrt{q}+224 q+980 q^{3/2}+3472 q^2+\dots,\\
\chi_{\frac23}^{\TF_7}&= \chi_{\frac{2}{3},w_2}^{SO(8)_3} +\chi_{\frac76,w_1+w_2}^{SO(8)_3}= q^{\frac{5}{18}}(28+160 \sqrt{q}+784 q+3080 q^{3/2}+\dots)  ,\\
\chi_{\frac56}^{\TF_7}&=  \chi_{\frac{5}{6},w_3+w_4}^{SO(8)_3} +\chi_{\frac43,w_1+w_3+w_4}^{SO(8)_3}=  q^{\frac59}(56+350 \sqrt{q}+1568 q+5704 q^{3/2}+\dots)  ,\\
\chi_{\frac89}^{\TF_7}&= \chi_{\frac89,2w_3}^{SO(8)_3}+\chi_{\frac{25}{18},w_1+2w_3}^{SO(8)_3} =35 \sqrt{q}+224 q+980 q^{3/2}+3472 q^2+\dots .
\ea
\ee
Clearly, this $\Gamma_{\theta}$ modular invariant of WZW $SO(8)_3$ is induced by the simple current $v$ that exchanges the affine node and vector node of affine $SO(8)$. The degeneracy two of the weight-$9/8$ fields is natural since there are the symmetry between the spinor and conjugate spinor nodes. The relations between affine characters and  fermionic characters including other sectors  can also be found in \cite[Equation (3.31)]{Bae:2021lvk}. The $\TF_{7}$ and $\TF_{11}$ images should form a holomorphic SCFT of $c=24$. The NS characters satisfy the bilinear relation $\TF_{7}\cdot \TF_{11}=K(\tau)^2-384$. 
Similarly, we find $\TF_{1}\cdot \TF_{17}=K(\tau)^2+K(\tau)-516$. 
\begin{table}[ht]
\def\arraystretch{1.1}
	\centering
	\begin{tabular}{|c|c|c|c|c|c|c|c|c|c|c|c|c|c|}
		\hline
		 $c$ & $h_{\rm NS}$  & $m_{1/2}$ & remark  &$\tc$ & $\tilh_{\rm NS}$ &  $\tm_{1/2}$ & remark  \\
		\hline
$\frac{4}{3}$ & $ \frac{1}{18},\frac{1}{6},\frac{1}{3},(\frac59)_2 $ & $ 1 $  &  $\TF_1,SM_{\rm sub}(36,2) $ & $\frac{68}{3}$ & $ \frac{17}{18},\frac{7}{6},\frac{4}{3},(\frac{13}{9})_2 $ & $ 0 $  &  $\TF_{17} $  \\
$\frac{20}{3}$ & $ \frac{5}{18},\frac{1}{3},\frac{2}{3},(\frac{7}{9})_2 $ & $ 10 $  &  $\TF_5,\star $ & $\frac{52}{3}$ & $  \frac{13}{18},\frac{5}{6},\frac{7}{6},(\frac{11}{9})_2$ & $0  $  &  $\TF_{13},\star $   \\
$\frac{28}{3}$ & $ \frac{7}{18},\frac{2}{3},\frac{5}{6},(\frac{8}{9})_2 $ & $  0$  &  $\TF_7 ,\cF(D_4)_3$ & $\frac{44}{3}$ & $ \frac{11}{18},\frac{2}{3},\frac{5}{6},(\frac{10}{9})_2 $ & $ 0 $  &  $\TF_{11} $   \\
		\hline
		\end{tabular}
			\caption{Fermionic Hecke images of $SM_{\rm sub}(36,2)$.}
			\label{tb:SM362sub}
		\end{table}

\section{Comments on $\rm\widetilde{NS}$ and R characters}\label{sec:R}
Although the most natural setting for fermionic Hecke operator is in the NS sector, one can still generalize it to the  $\rm\widetilde{NS}$ and R sectors by some trick. As we  reviewed in Section \ref{sec:basic}, under $S$ action, the $\rm\widetilde{NS}$ and R sectors transform into each other. Therefore, it is necessary to consider them together, which we denote as the $(\rm\widetilde{NS},R)$ sector. The $T^2$ and $S$ actions are now closed in this combined sector. Denote the combined character as $\chi_{(\rm\widetilde{NS},R)}=\{\chiNSt,\chiR\}$. Suppose there are $n$ number of NS characters, then there exists a $2n$-dimensional representation $\rho^{\rm F'}:\Gamma_\theta\to GL(2n,\IC)$ such that for any $\gamma\in\Gamma_\theta$, 
\be
\chi_i^{(\rm\widetilde{NS},R)}(\gamma \tau)=\sum_{j}\rho^{\rm F'}(\gamma)_{ij}\chi_j^{(\rm\widetilde{NS},R)}(\tau),\qquad i=1,2,3,\dots,2n.
\ee
To define the fermionic Hecke operator for $(\rm\widetilde{NS},R)$ sector, we still need the transfer matrix $\rho^{\rm F'}(\sigma_p)$. We find it can be defined by   $\rho^{\rm F'}(S)$ and $\rho^{\rm F'}(T^2)$ just analogous to \eqref{eq:sigmaF} as 
\be\label{sigmapR}
\rho^{\rm F'}(\sigma_p) =\rho^{\rm F'}\Big( S^2(T^2)^{\frac{\bar{p}^2-\bar{p}}{2}}S (T^2)^{-\frac{{p}-1}{2}} S(T^2S)^{\bar{p}-1}\Big).
\ee
Then the fermionic Hecke operator $\TF_p$ is defined in the same way as \eqref{Tp} and all properties of Hecke images still hold. 
One prominent feature here is that $\TF_p$ always maps the $(\rm\widetilde{NS},R)$ characters of one theory to the $(\pm\rm\widetilde{NS},R)$ characters of another theory. The reason is that there are always even numbers of $S$ actions in the right hand side of \eqref{sigmapR}, which protects the order. 
We will show three examples here: one chiral fermion, $(SLY)_1$ and $SM(5,3)$. Other types are just similar. 

For one single chiral fermion, it is well-known that the $\rm\widetilde{NS}$ and R characters are $\psi_{\rm \widetilde{NS}}=\sqrt{{\theta_4}/{\eta}}$ and $\psi_{\rm R}=\sqrt{{\theta_2}/{\eta}}.$ Under $S$ action, the $\rm\widetilde{NS}$ and R characters are exchanged. Thus 
\be
\rho^{\rm F'}(S)=\left(
\begin{array}{cc}
 0 & 1 \\
 1 & 0 \\
\end{array}
\right),\qquad \rho^{\rm F'}(T^2)=\left(
\begin{array}{cc}
 e^{-\frac{i \pi}{12} } & 0 \\
 0 & e^{\frac{i \pi }{6}} \\
\end{array}
\right) .
\ee
By \eqref{sigmapR} we find there exist two classes of $\rho^{\rm F'}(\sigma_p)$: for $p=1, 7, 17, 23\textrm{ mod } 24$, i.e., $p^2\equiv 1 \textrm{ mod } 48$, $
\rho^{\rm F'}(\sigma_p)=\mathrm{Id},$ 
while for $p=5, 11, 13, 19 \textrm{ mod } 24$, i.e., $p^2\equiv 25 \textrm{ mod } 48$, $\rho^{\rm F'}(\sigma_p)=-\mathrm{Id}$. For fermionic Hecke images with $p<24$, we find that for $p=1, 7, 17, 23$, $\TF_p$ describes the $\{\chiNSt,\chiR\}$ of fermionization $\cF(SO(p)_1)$, while for $p=5, 11, 13, 19$, it describes the $\{-\chiNSt,\chiR\}$ of fermionization $\cF(SO(p)_1)$.  Here we adopt the usual convention that the leading Fourier coefficient of vacuum $\chiNSt$  is 1. It is easy to check the bilinear relation $\TF_p\cdot\TF_{24-p}=K(\tau)+24$, which resembles the bilinear relation in the NS sector but has different meaning.

For supersymmetric Lee-Yang model $(SLY)_1$, the R weights are $h_{\rm R}=\frac{1}{32},\frac{5}{32}$. The $\rm\widetilde{NS}$ and R characters can be defined from the bosonic minimal model $M_{\rm eff}(13,2)$ by $\{\chiNSt,\chiR\}=\{\chi_0-\chi_{1/2},\chi_{1/4}-\chi_{7/4},\chi_{1/32}+\chi_{33/32},\sqrt{2}\chi_{5/32}\}$. See the expression of $M_{\rm eff}(13,2)$ characters in e.g. \cite[Equation (8.3)]{Duan:2022ltz}.  It is worthy to point out that the $\rm\widetilde{R}$ character here is just a constant $\chi_{1/32}-\chi_{33/32}=1$. We find the $S$ and $T^2$ matrices for the $\{\chiNSt,\chiR\}$ are
\be
\rho^{\rm F'}(S)=\frac{1}{\sqrt{2}} \left(
\begin{array}{cccc}
 0 & 0 & 1 & 1 \\
 0 & 0 & 1 & -1 \\
 1 & 1 & 0 & 0 \\
 1 & -1 & 0 & 0 \\
\end{array}
\right),\qquad \rho^{\rm F'}(T^2)=\left(
\begin{array}{cccc}
 e^{-\frac{i \pi}{8} } & 0& 0& 0 \\
  0& e^{\frac{7i \pi }{8}} & 0& 0\\
 0  & 0 &1 & 0  \\
 0  & 0  & 0 & e^{\frac{i \pi}{2} }\\
\end{array}
\right) .
\ee
Consider the fermionic Hecke operation on  $\{\chiNSt,\chiR\}$. We find there are in total two fermionic Hecke classes: for $p\equiv 1, 7 \!\!\mod 8$, $\rho^{\rm F'}(\sigma_{p})=\mathrm{Id}$, for $p\equiv 3,5  \!\!\mod 8$,
\be
\rho^{\rm F'}(\sigma_{p})=  \left(
\begin{array}{cccc}
 0 & -1 & 0 & 0 \\
 -1 & 0 & 0 & 0 \\
 0 & 0 & -1 & 0 \\
 0 & 0 & 0 & 1 \\
\end{array}
\right).
\ee
We explicitly compute all $\TF_p$ for $p<16$ and find that the $\TF_3$ image exactly describes the $\{-\chiNSt,\chiR\}$ of $\cF(A_1)_6$ theory given in \eqref{SU26allchi}, while the $\TF_{13}$ image exactly describes the $\{-\chiNSt,\chiR\}$ of $\cF(C_6)_1$ theory. It is easy to check the bilinear relation $\TF_3\cdot\TF_{13}=K(\tau)+24$.

For unitary supersymmetric minimal model $SM(5,3)$, the R weights are $h_{\rm R}=\frac{3}{80},\frac{7}{16}$. The $\rm\widetilde{NS}$ and R characters can be defined from the bosonic tricritical Ising model $M(5,4)$ by $\{\chiNSt,\chiR\}=\{\chi_0-\chi_{3/2},\chi_{1/10}-\chi_{3/5},\sqrt{2}\chi_{3/80},\sqrt{2}\chi_{7/16}\}$. The $S$ matrix here is related to the $\rho^{\rm F}(S)$ matrix of the NS characters of $SM_{\rm sub}(60,2)$ in \eqref{SSM602} by a simple relation
\be
\rho^{\rm F'}(S)= \left(
\begin{array}{cc}
 0 & \rho^{\rm F}(S) \\
 \rho^{\rm F}(S) & 0 \\
\end{array}
\right) .
\ee
Consider the fermionic Hecke operation on $\{\chiNSt,\chiR\}$. We find there exist four Hecke classes: for $p\equiv 1, 11, 29, 41,$ $79, 91, 109, 119 \!\!\mod 120$, $\rho^{\rm F'}(\sigma_{p})=\mathrm{Id}$, for $p\equiv 19, 31, 49, 59, 61,$ $71, 89, 101  \!\!\mod 120$, $\rho^{\rm F'}(\sigma_{p})=-\mathrm{Id}$,
    \be
    \ba
&\textrm{for }  p\equiv 7, 37, 43, 47, 73, 77, 83, 113\!\!\!\mod 120,\quad\  
\rho^{\rm F'}(\sigma_{p})=\left(
\begin{array}{cccc}
 0 & -1 & 0 & 0 \\
 1 & 0 & 0 & 0 \\
 0 & 0 & 0 & -1 \\
 0 & 0 & 1 & 0 \\
\end{array}
\right) ,\\ 
&\textrm{for }  p\equiv 13, 17, 23, 53, 67, 97, 103, 107  \!\!\!\mod 120,\quad 
\rho^{\rm F'}(\sigma_{p})= \left(
\begin{array}{cccc}
 0 & 1 & 0 & 0 \\
 -1 & 0 & 0 & 0 \\
 0 & 0 & 0 & 1 \\
 0 & 0 & -1 & 0 \\
\end{array}
\right).
\ea
\ee 
These four fermionic Hecke classes form a $\IZ_4$ group. However, we remark that these are different from the four Hecke classes of the NS sector discussed in Section \ref{sec:SM53}. By explicitly computing the fermionic Hecke images, we find that the $\TF_{19}$ image describes the $\{-\chiNSt,\chiR\}$ characters of fermionization $\cF(E_7)_2$. We refer the relations between $\{\chiNSt,\chiR\}$ and affine $(E_7)_2$ characters to e.g. \cite{Bae:2020xzl}.

\section{Summary and outlook}\label{sec:outlook}
This work gives an affirmative answer to the question raised in the end of \cite{Duan:2022ltz} whether there exists a natural fermionic Hecke operator for  2d fermionic RCFTs. We find the natural setting is the NS characters of fermionic RCFTs, as they transform to themselves by $S$ and $T^2$ actions. In a mathematical sense, we find the natural definition of Hecke operator for vector-valued $\Gamma_\theta$ modular forms of weight zero. We also discuss how to apply fermionic Hecke operator to the $\rm\widetilde{NS}$ and R sectors, that is to combine them together such that the characters $\{\chiNSt,\chiR\}$ still form a vector-valued $\Gamma_\theta$ modular form.

We discover many fermionic Hecke relations among the 2d fermionic and supersymmetric RCFTs, most of which are unknown from the bosonic side. In particular, for all supersymmetric theories appeared in \cite{Johnson-Freyd:2019wgb,Bae:2020xzl,Bae:2021lvk,Bae:2021mej}, the only one we couldn't find a proper realization as a fermionic Hecke image is the fermionization of $ SU(8)_2/\mathbb{Z}_2$ theory \cite[Equation (4.5)]{Bae:2021lvk}, with central charge $c=\frac{63}{5}$ and NS weights $h_{\rm NS}=\frac45,(\frac{9}{10})_2$. The weights and degeneracy reminds us of $SM(5,3)^2$ with $c=\frac{7}{5}$ and NS weights $h_{\rm NS}=(\frac{1}{10})_2,\frac15$ which was briefly mentioned in Section \ref{sec:SM53}. One can easily check a $\TF_9$ (if exists) on $SM(5,3)^2$ would produce the correct degeneracy and multiple $9$ condition for the central charge and NS weights. Unfortunately,  9 is not coprime to the conductor $120$ of $SM(5,3)^2$. It would be interesting to determine whether there exists a middle fermionic theory serving as the generalized $\TF_3$ image with $c=\frac{21}{5}$ and conductor $N=40$. Such theory should be non-unitary according to the classification of rank-8 SMC.

The fermionic Hecke operator also helps us to rule out some previously undetermined theory. For example, by studying  the third order fermionic MLDEs, \cite{Bae:2021mej} bootstraped the three NS characters of a potential $c=\frac{66}{5}$ theory. As we discussed in Section \ref{SMsub202}, we notice this theory can be realized a $\TF_{11}$ image of $SM_{\rm sub}(20,2)$ and share the same $S$-matrix and fusion rules as $SM_{\rm sub}(20,2)$. Therefore, it is non-unitary and can not be the fermionization of any WZW models.

The relation between fermionic Hecke operator and super modular category should be further clarified. It should be possible to generalize the results on Galois symmetry in \cite{Harvey:2019qzs} to the fermionic cases.  It is also intriguing to consider whether fermionic Hecke operation can produce the fermionic characters of the new putative objects in the rank-10 SMC recently proposed in \cite{Cho:2022kzf}. We leave these for future work.

\section*{Acknowledgements}
We would like to thank Gil Young Cho, Zhihao Duan, Hee-Cheol Kim, Sungjay Lee, Linfeng Li, Minyoung You and Haowu Wang for useful discussions. KL and KS are supported by KIAS Grants PG006904 and QP081001 respectively. KL is also supported in part by the National Research Foundation of Korea (NRF) Grant funded by the Korea government (MSIT) (No.2017R1D1A1B06034369).

\end{document}